\documentclass[sigconf]{acmart}

\usepackage{xspace}
\usepackage{balance}
\AtBeginDocument{%
  }
    
\setcopyright{acmcopyright}
\copyrightyear{2025}
\acmYear{2025}
\acmDOI{XXXXXXX.XXXXXXX}

\acmConference[CHI '25]{ACM SIGCHI Conference on Human Factors in Computing Systems (CHI)}{April, 2025}{Japan, Yokohama}
\acmPrice{15.00}
\acmISBN{978-1-4503-XXXX-X/18/06}

\begin{document}

\newcommand{\papername}{PinchCatcher\xspace}
\newcommand{\tit}{\papername: Enabling Multi-selection for Gaze+Pinch\xspace}
\title{\tit}

\author{Jinwook Kim}
\email{jinwook.kim31@kaist.ac.kr}
\affiliation{%
  \institution{Graduate School of Culture Technology}
  \institution{KAIST}
  \city{Daejeon}
  \country{South Korea}}

\author{Sangmin Park}
\email{psmdc0714@kaist.ac.kr}
\affiliation{%
  \institution{Graduate School of Culture Technology}
  \institution{KAIST}
  \city{Daejeon}
  \country{South Korea}}
  
\author{Qiushi Zhou}
\email{qiushi.zhou@cs.au.dk}
\affiliation{%
  \institution{Aarhus University}
  \city{Aarhus}
  \country{Denmark}}

\author{Mar Gonzalez-Franco}
\email{margonzalezfranco@gmail.com}
\affiliation{%
  \institution{Google}
  \city{Seattle, WA}
  \country{USA}}
  
\author{Jeongmi Lee}
\authornote{Co-corresponding author}
\email{jeongmi@kaist.ac.kr}
\affiliation{%
  \institution{Graduate School of Culture Technology}
  \institution{KAIST}
  \city{Daejeon}
  \country{South Korea}}
  
\author{Ken Pfeuffer}
\authornotemark[1]
\email{ken@cs.au.dk}
\affiliation{%
  \institution{Aarhus University}
  \city{Aarhus}
  \country{Denmark}}

\renewcommand{\shortauthors}{Kim et al.}

\begin{abstract}
This paper investigates multi-selection in XR interfaces based on eye and hand interaction. We propose enabling multi-selection using different variations of techniques that combine gaze with a semi-pinch gesture, allowing users to select multiple objects, while on the way to a full-pinch. While our exploration is based on the semi-pinch mode for activating a quasi-mode, we explore four methods for confirming subselections in multi-selection mode, varying in effort and complexity: dwell-time (SemiDwell), swipe (SemiSwipe), tilt (SemiTilt), and non-dominant hand input (SemiNDH), and compare them to a baseline technique. In the user study, we evaluate their effectiveness in reducing task completion time, errors, and effort. The results indicate the strengths and weaknesses of each technique, with SemiSwipe and SemiDwell as the most preferred methods by participants. We also demonstrate their utility in file managing and RTS gaming application scenarios. This study provides valuable insights to advance 3D input systems in XR.
\end{abstract}



\begin{teaserfigure}
  \includegraphics[width=\textwidth]{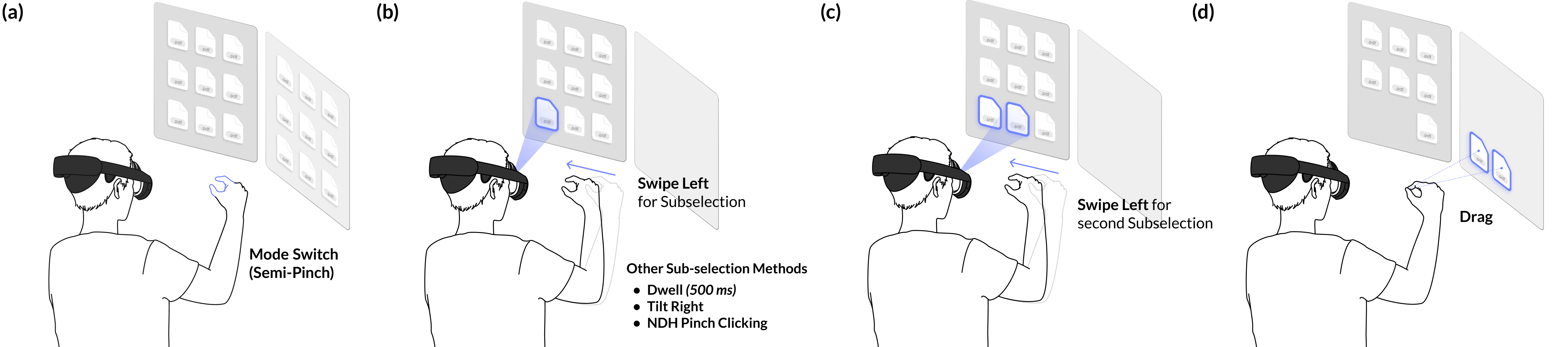}
  \caption{We simplify multi-selection for eye and hand interfaces in extended reality (XR) by utilizing the pre-pinch phase (a) —when users initiate but do not complete a pinch gesture— to contextualize upcoming actions. In this quasi-mode, users can gaze at multiple objects, subselecting them for the subsequent full-pinch command (b-c), allowing for intuitive, pre-pinch multi-selection before initiating gesture operations like drag and drop (d).}
  \label{fig:teaser}
  \Description{We simplify multi-selection for eye and hand interfaces in extended reality (XR) by utilizing the pre-pinch phase—when users initiate but do not complete a pinch gesture—to contextualize upcoming actions. In this quasi-mode, users can gaze at multiple objects, subselecting them for the subsequent full-pinch command, allowing for intuitive, pre-pinch multi-selection before finalizing gestures. When users maintain a semi-pinch posture (a), the system interprets this signal as a mode-switch to a pre-pinch state, which allows them to contextualize the subsequent full gesture. In this state,  users can select multiple objects by simply gazing at each object of interest (b-c). When complete, a pinch-contact gesture manipulates all objects at once (d).}
\end{teaserfigure}

\maketitle
\section{Introduction}
The gaze + pinch has the potential to become as widely available in virtual reality (VR) as touch input in tablets and smartphones, as a direct natural interaction~\cite{pfeuffer2024design, pfeuffer2017gazeP}. Still, it has several shortcomings, like the ability for preshaping of our hand~\cite{lystbaek2024hands}, competing attentional demands of gaze in augmented reality (AR)~\cite{pfeuffer21}, and a limited throughput in multi-selection scenarios. On smartphones, we have moved away from having to tap on every item or photo in our gallery to select fast, by adding hold-and-swipe type of gestures. We need our equivalent for gaze + pinch.

Multi-selection techniques render the process of selecting multiple items such as files or images more efficient~\cite{lucas2005design, wu2023point, xia2017collection}. Multi-selection can be enabled through a \textit{persistent} mode, such as long-pressing on a touchscreen, that stays active until manually turned off~\cite{raskin2000humane}. Alternatively, a \textit{quasi-mode} (or spring loaded~\cite{hinckley2006springboard}), such as when holding the CTRL key in a mouse and keyboard-based UI, which is suitable for spontaneous multi-selections.

Although multi-selection has matured for traditional computing devices, it is open to how eye- and hand-tracked UIs in extended reality (XR) can support such a mechanism. Many XR UIs are based on the canonical pinch gesture, offering dexterous and natural ways to manipulate single objects. Multi-selection is supported via persistent modes (e.g., special gestures like long-pinch or dedicated UI buttons~\cite{blocks, shapesxr, fernandez2022multi, lucas2005design}), but quasi-mode-based multi-selection features are difficult to integrate. To date, only Shi et al. have explored this area in the context of gesture-only UIs~\cite{shi2024experimental}. Their approach introduces additional gestures that indeed enable mode-switching but it trades it with higher effort as users need additional cognitive resources to learn and remember them \cite{shi2024experimental, norman10}. We propose a different approach by designing for multi-selection without departing from the intuitive pinch-based input paradigm.


People naturally look at objects of interest in point-select tasks environments~\cite{zhai1999manual, Liebling14, piumsomboon17, turkmen2024eyeguide}. Modern headsets (e.g., Hololens 2, Quest Pro, Vision Pro) exploit this behavior, offering a multimodal UI based on user inputs interpreted through gaze and pinch gestures \cite{pfeuffer2017gazeP, Chatterjee15, velloso2015empirical, wagner2023fitts}--we posit that this behavior will hold specifically in the context of manipulating multiple objects at once. Before the manipulation commences, users are likely to look at the objects they want to manipulate, giving away which and how many targets are to be manipulated. As well, users are likely to prepare their pinch gesture in anticipation of the subsequent manipulation~\cite{zhu2023pinchlens}. Taken together, we experiment with this dual phenomenon for implicitly specifying all objects of interest with gaze just before one engages in a pinch gesture.

We propose \papername, a new approach for multi-selection that allows subselecting multiple targets for single pinch gestures in eye- and hand-tracked XR. The idea is to integrate gaze pointing with a semi-pinch gesture \cite{zhu2023pinchlens} that acts as a quasi-mode. This state occurs when users maintain their fingers in a partially pinched position, hovering between fully extended and fully flexed. Figure~\ref{fig:teaser} illustrates the concept. When users maintain a semi-pinch posture (a), the system interprets this signal as a mode-switch to a pre-pinch state, which allows them to contextualize the subsequent full gesture. In this state,  users can select multiple objects by simply gazing at each object of interest (b-c). When complete, a pinch-contact gesture manipulates all objects at once (d).

This leads us to our main research question: how can users, while holding a semi-pinch state, add more targets to the selection by gaze without accidentally including unwanted items (i.e., Midas Touch problem~\cite{jacob1990you, penkar2012designing, isomoto2022interaction})? As users are only briefly holding a semi-pinch gesture to contextualize their subsequent action, it needs to balance simplicity of input with low cognitive and physical effort, all while being sufficiently robust to clarify intent. We explore four distinct methods for triggering subselection while users maintain the semi-pinch state, using gaze as the primary pointer: (1) SemiDwell, a low-effort technique based on dwell time with the eyes alone; (2) SemiSwipe and (3) SemiTilt, both utilizing gestures from the same hand; and (4) SemiNDH, where the non-dominant hand (NDH) performs pinch gestures while the dominant hand (DH) maintains the semi-pinch.

We present a user study that compares the four techniques against a baseline (FullDH) that closely resembles the CTRL key approach of desktop UI~\cite{jarvi2016one, wills1996selection}. Here, the user holds a pinch gesture of the NDH, and normal pinch gestures of the DH will add to the multi-selection. These multi-selection techniques are tested in a serial selection task~\cite{wu2023point, bergstrom2021evaluate} where users select each object individually. This task examines fundamental subselection capabilities, which are critical for guiding future developments, such as area selection~\cite{kim2018tpvr, zhang2023multi, shi2023exploring}. Objects were arranged in a 2D grid in space, reflecting common layouts in current XR UIs. Participants were tasked with selecting 2, 4, or 6 targets, focusing on our quasi-mode techniques, which are intended for scenarios involving a few targets, rather than a large number where persistent modes might be more effective. Based on the findings, we demonstrate the idea of using \papername in two applications: file management, which is a fundamental task when using the computer, and real-time strategy (RTS) game for usage in the 3D environment. 

The results of the study show the following findings:

\begin{itemize}
    \item There was no significant difference in task completion time across the techniques.
    \item Participants selected more distractors with the FullDH technique than all other techniques but could correct them before finalizing the selection, leading to the lowest error rate.
    \item The SemiSwipe technique resulted in fewer distractor selections than SemiDwell and SemiNDH, and overall fewer errors than SemiDwell and SemiTilt.
    \item SemiNDH led to more perceived physical effort than SemiDwell and FullDH.
    \item Most participants preferred SemiDwell (9) and SemiSwipe (8).
\end{itemize}

In summary, our contributions are as follows: First, we introduce \papername, an interaction concept for multi-selection in XR that combines the benefits of a semi-pinch quasi-mode with rapid eye movement-based input. Second, we propose four distinct methods for confirming subselections during the semi-pinch state, each with its strengths and limitations. Third, we present a user study that provides insights into user performance and experience, demonstrating the effectiveness of our techniques compared to a bimanual approach. Fourth, we offer application probes, such as photo management and RTS gaming, to showcase use cases and provide guidance for implementing \papername. Our work provides valuable insights for future 3D input solutions and highlights o new opportunities to advance eye- and hand-tracked interaction in XR.


\section{Related Work}
\subsection{Multi-selection Interaction in Representative Platforms}
Multi-selection especially provides convenience for performing the same task on multiple targets (e.g., dragging, copying, and deleting)~\cite{lucas2005design}. The multi-selection consists of two steps: mode switching and triggering subselection. Users first switch to multi-selection mode from single selection mode. Then, they target and confirm the grouping as they did in a single selection. Additionally, there are serial and parallel ways of grouping~\cite{shi2023exploration, lucas2005design}. Serial grouping is a method that selects target objects individually, while parallel methods group multiple targets simultaneously. For the serial way, users individually select the targets with a basic selection input method such as a mouse click, direct touch, and ray interaction for grouping~\cite{wills1996selection, wu2023point}. In terms of the parallel way, users utilized the predefined 3D cones or spotlights shape region~\cite{lucas2005design, steed2006towards, shi2024experimental} or generate a region by selecting two points with a lasso or a selection box tool for grouping~\cite{shi2023exploring, zhang2023multi, wu2023point}.

\begin{figure*}
  \includegraphics[width=\textwidth]{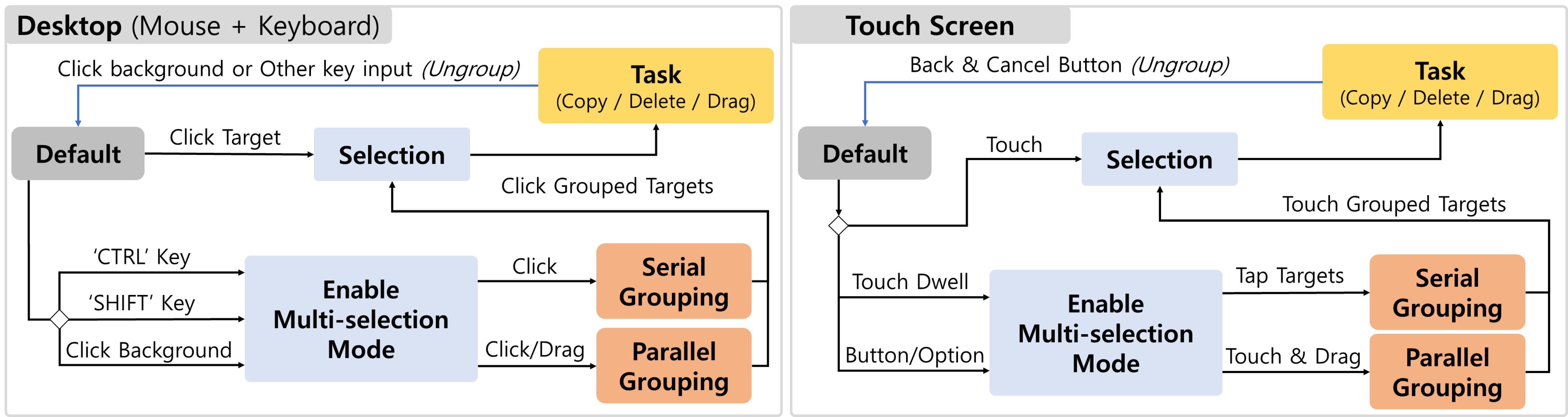}
  \caption{A diagram of a multi-selection process in 2D desktop and touch screen environments.}
  \label{fig:MOSDiagram}
  \Description{A diagram of a multi-selection process in 2D desktop and touch screen environments. In the desktop environment, users perform multi-selection by combining keyboard and mouse inputs. For instance, users press the CTRL or SHIFT button on the keyboard to activate multi-selection as a quasi-mode and group targets with the mouse or arrow button input. Regarding multi-selection in the touch screen, users touch one of the components and dwell on activating the persistent multi-selection mode or enabling it with the mode switching button UI.}
\end{figure*}

In the desktop environment (Figure~\ref{fig:MOSDiagram}), users perform multi-selection by combining keyboard and mouse inputs~\cite{jarvi2016one, wills1996selection}. For instance, users press the CTRL or SHIFT key on the keyboard to activate multi-selection as a quasi-mode and group targets with the mouse or arrow button input. Regarding multi-selection in the touch screen (Figure~\ref{fig:MOSDiagram}), users touch one of the components and dwell on activating the persistent multi-selection mode or enabling it with the mode switching button UI~\cite{brandl2008combining, pfeuffer2014gaze, ramanathan2016multi, dehmeshki2010design}.

Overall, two types of multi-selection activation methods have been utilized in the previous platforms: a quasi-mode that maintains the input for the mode switch while grouping, and a persistent mode that separates the mode switch and grouping process. Users may lack awareness of the selection status in the persistent mode state, leading to unintended selection with different modes. To illustrate, users could attempt to perform a multi-selection but were inadvertently in a single selection mode. Consequently, they would have to change the mode and perform the multi-selection from the beginning. In contrast, since users must maintain their actions in quasi-mode-based techniques, it could prevent potential errors by supporting their awareness of the selection status~\cite{hinckley2006springboard}.

Similarly, a VR environment with a physical controller uses both types of methods by using physical button/joystick inputs or menu UI to change the selection mode~\cite{wu2023point, fernandez2022multi, zhao2023metacast}. In terms of direct physical input, Wu et al. explored the performance and usability of the various multi-selection methods, directly touching each object with the controller serially or making an arbitrary area and modifying the size with the controller movement for parallel selection~\cite{wu2023point}. The menu UI for enabling multi-selection mode is utilized in the recent VR and AR applications (i.e., ShapesXR, Google Blocks)~\cite{blocks, shapesxr}. For instance, in the ShapeXR application, the user has to press the button on the controller to activate the pop-up menu and select the multi-selection option with the ray-casting from the controller. Then they select which multi-selection technique to use (i.e., serial group selection, lasso area selection) and begin grouping the targets.

Several studies have proposed techniques that utilized additional interfaces (i.e., virtual or physical tablets) with the physical controller to assist multi-selection in 3D environments~\cite{lucas2005design, montano2020slicing}. For instance, Lucas et al. proposed a method that uses a virtual tablet that shows a camera view image of a specific point and can draw an area on the image for multi-selection~\cite{lucas2005design}. There are also techniques based on the metaphor of actions from daily life~\cite{huang2019review, li2022designing}. Li et al. proposed a multi-selection technique based on sewing action with the controller~\cite{li2022designing}. The user evaluation results showed that these techniques had improved the performance of multi-selection in specific contexts (i.e., typing, 3D modeling). However, it requires multiple button inputs to change the selection mode and enable the multi-selection technique. This process could slow down the performance of grouping targets and switching between modes, which makes it hard for this technique to be used in broader XR contexts, such as game and remote co-working applications.

\subsection{Multi-selection for Hand and Eye Interaction in XR}
Gaze input enables fast, natural target acquisition~\cite{chatterjee2015gaze, jacob1991use}, while hand gestures are highly suitable for expressive commands~\cite{Wobbrock09, bragdon2011gesture, hyrskykari2012gaze}. Their multimodal fusion—where the eyes target and the hands 'click'—has shown great potential across various human-computer interaction paradigms, including mouse-operated systems~\cite{zhai1999manual, jacob1990you}, touch devices~\cite{pfeuffer2014gaze, Stellmach12}, pen computing~\cite{Pfeuffer15}, and 3D gesture-controlled desktops~\cite{chatterjee2015gaze, velloso2015empirical}. Modern XR Head Mounted Displays (HMDs) increasingly embed eye tracking sensors which have numerous applications~\cite{plopski2022eye}, such as for tasks like one-handed menu control~\cite{pfeuffer2023palmgazer}, 3D sketching~\cite{Turkmen24}, and hands-free accessibility~\cite{Sidenmark23}. A key interaction model, gaze + pinch, follows carefully crafted design principles that could be utilized in universal XR operating systems~\cite {pfeuffer2024design, pfeuffer2017gazeP}, as exemplified by the Vision Pro HMD.

Accordingly, we build upon this interaction model by experimenting with multi-selection techniques. Studies have proposed multi-selection techniques for the hand-based interaction contexts~\cite{zhang2023multi, kim2018tpvr, lee2016tunnelslice, shi2024experimental}, by using both hand pinch gesture~\cite{kim2018tpvr} or ray from multiple finger~\cite{zhang2023multi} to make an arbitrary area for group selection. Complementary to the work on multi-selection, Surale et al. explored mode-switching through various gestures such as a fist, palm, pinching, and pointing in both DH and NDH~\cite{surale2019experimental}. They concluded that the pinch with DH is a viable option for a controller button for mode-switching, which inspired us to take a deeper look at this gesture. Hu et al. investigated using gaze input for mode-switching on a pen and tablet device \cite{hu2023gaze}. They found that gaze alone can provide an efficient mode-switch with low effort, however, a multimodal variant where the user fixated on a distinct UI element coupled with a manual trigger led to more robust mode-switches~\cite{hu2023gaze}. Inspired by this, our interaction design considers how gaze and pinch gestures can be used together for a mode switch to a multi-selection mode.

Multi-selection techniques that combine eye gaze with hand gestures are rarely considered in the literature. For instance, a closely related work is Shi et al. work on region selection where the user defines two points that form the diagonal of a region~\cite{shi2023exploring}, which in theory can be used for area selection. They evaluated four techniques: Gaze\&Finger, Gaze\&Pinch, Pinch, and Eyeblink, and found that both unimodal approaches work best. Our work is complementary, as we consider a new way of semi-pinch as a mode-switch, to mode a pinch gesture. This affords flexible gestures without altering the default operation of pinch-click, drag, and so on, and avoids overlap with the gestures employed for other functions~\cite{benko2009beyond, pfeuffer2024design}. Further, we do not consider eyes-only approaches as a mode-switch, as the active use of the eyes for UI manipulations alters the natural division of labor between the eye and hand modalities \cite{pfeuffer2024design}. As well, if the interaction relies too much on the gaze, it could cause eye fatigue and errors~\cite{Hirzle22}. When designing interaction with gaze and hand, it is essential to design for the natural roles of each modality and exploit their synergies~\cite{kim2023exploration, zhai1999manual}-- which is here, that users naturally focus on the candidates for multi-selection, coupled with an explicit pre-pinch event.



\section{Design of Serial Multi-selection Techniques for Gaze + Pinch Interaction}
We describe the design of multi-selection techniques for eye-hand interaction systems that take natural gaze behavior into account and that are compatible with the intuitive pinch-based manipulation paradigm. Here we first lay out our design rationale for choosing serial multi-selection and one-handed interaction. We then lay out our four multi-selection techniques. We also describe a mouse and keyboard-inspired pinch technique (FullDH) as a baseline for comparison to the semi-pinch-based techniques.

\subsection{Design Considerations}
\subsubsection{Serial Way of Multi-selection}
Among the two ways of multi-selection (i.e., serial and parallel), the parallel way seems to be faster since it is capable of grouping multiple targets with one action. However, previous studies have proposed that the performance and usability of serial and parallel ways significantly differ depending on the context~\cite{wu2023point, lucas2005design}. For instance, if a large number of targets are distributed close to each other, the parallel methods have an advantage over the serial methods. On the other hand, if the number of targets is small (below 10) and targets are situated near distractors, the serial methods demonstrate enhanced performance. Both methods are important for an efficient and comfortable multi-selection experience, but parallel methods require structured environments that are less general in the XR context. Thus, we focused the scope of the current study on the serial grouping methods.

\subsubsection{Maintain One-hand Interaction for Gaze + Pinch}
In terms of designing a multi-selection method that could maximize the advantage of gaze, it is essential to consider the connectivity with the functions (e.g., selection, dragging, and resizing) that were performed with the Gaze + Pinch previously (cf. `\textit{Flexible Gesture}'~\cite{pfeuffer2024design}). For instance, a single pinch gesture with the DH is intuitively assigned to a single object selection, and a pinch with both hands for resizing or rotating a selected object. Adding a new hand gesture instead of a pinch might require users to learn additional gestures or actions. In addition, using the other hand limits the freedom of the hand and the flexibility of other functions after or during multi-selection. Consider this example: if multiple selections are possible with a single hand, users can perform other tasks in parallel or tasks that could support the right-hand multiple selections (e.g., move objects that occlude target, scroll the list to search for more targets). Thus, we deliberately constrain our design to a one-handed interaction, retaining compatibility with the default interaction model of pinch-based manipulations.

\subsection{Semi-Pinch State for Mode Switching Method}
Based on the design considerations, we propose a mode-switching method using a semi-pinch gesture. The semi-pinch gesture was used in a previous study as a preselection state for displaying lens interface and adopting a control-display gain on hand to enable precise selection of small objects~\cite{zhu2023pinchlens}. Since the \textit{PinchLens} technique uses hand gestures for both selection and confirmation, it requires users to keep attention on tracking the pinch status (i.e., the distance between index and thumb fingertip) to manipulate the lens interface while conducting the task, and it is susceptible to the Heisenberg issues \cite{Wolf20}. In contrast, the Gaze + Pinch interaction separates these roles to "the eyes select, the hands manipulate"~\cite {pfeuffer2024design}, where the attention is not on the hand gesture and it is more robust to the Heisenberg issues~\cite{Wolf20}. When using semi-pinch and gaze, it must be clear to the user how to enter and exit the semi-pinch status without looking at the physical hand, as the eyes' attentional resources are reserved for the targets before finalizing a selection. 

\begin{figure}
  \includegraphics[width=\linewidth]{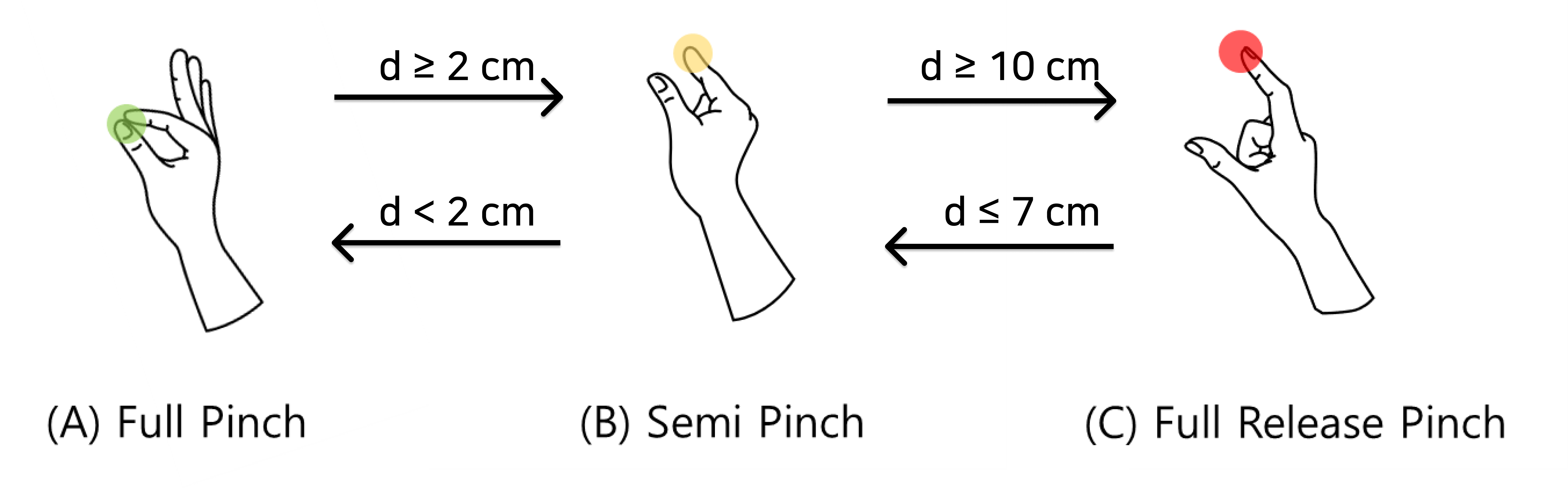}
  \caption{Illustration of the three hand-pinch states used to design the multi-selection method. Each state was detected based on the distance between the index and thumb fingertips. The indicator for each state is shown on the index fingertip. (A) Full pinch is detected when both fingertips are touching, and the indicator is highlighted in green. Users were able to interact with the selected objects in this state. (B) Semi-pinch is used to enable multi-selection mode and is activated when the fingertip distance is between 2 and 7 cm, as shown in the yellow indicator. (C) Full-release pinch is a state when the distance is over 10 cm and is indicated in red. It disables the grouping of all objects.}
  \label{fig:pinchStatusModel}
  \Description{Illustration of the three hand-pinch states used to design the multi-selection method. Each state was detected based on the distance between the index and thumb fingertips. The indicator for each state is shown on the index fingertip. (A) Full pinch is detected when both fingertips are touching, and the indicator is highlighted in green. Users were able to interact with the selected objects in this state. (B) Semi-pinch is used to enable multi-selection mode and is activated when the fingertip distance is between 2 and 7 cm, as shown in the yellow indicator. (C) Full-release pinch is a state when the distance is over 10 cm and is indicated in red. It disables the grouping of all objects.}
\end{figure}

As illustrated in Figure~\ref{fig:pinchStatusModel}, we mainly utilized three states of a pinch: full pinch, semi pinch, and full release pinch. When the full pinch is triggered, the grouping state ends, and users can start manipulating the grouped objects. In the semi-pinch state, the mode was changed to a multi-selection mode, and the full-release pinch made all grouped objects to an ungrouped state. We implemented the hand gesture detection using Meta Hand Pose Detection SDK and added the detection threshold between the full-release pinch and semi-pinch status to prevent the false activation of the ungrouping function. The distance between the index and thumb fingertips was used to implement the threshold, and a different value was applied for switching from each state to another state~\cite{velloso2017motion, zhu2023pinchlens}. Throughout the test, we selected an affordable distance for each threshold for the current study. As a result (Figure~\ref{fig:pinchStatusModel}), the transition from semi-pinch to full-release pinch states was activated when the fingertip distance reached 10 cm, while a semi-pinch state was activated when the distance was between 2 to 7 cm. These thresholds allow users to perform the pinch gestures more clearly.

\begin{figure*}
  \includegraphics[width=\textwidth]{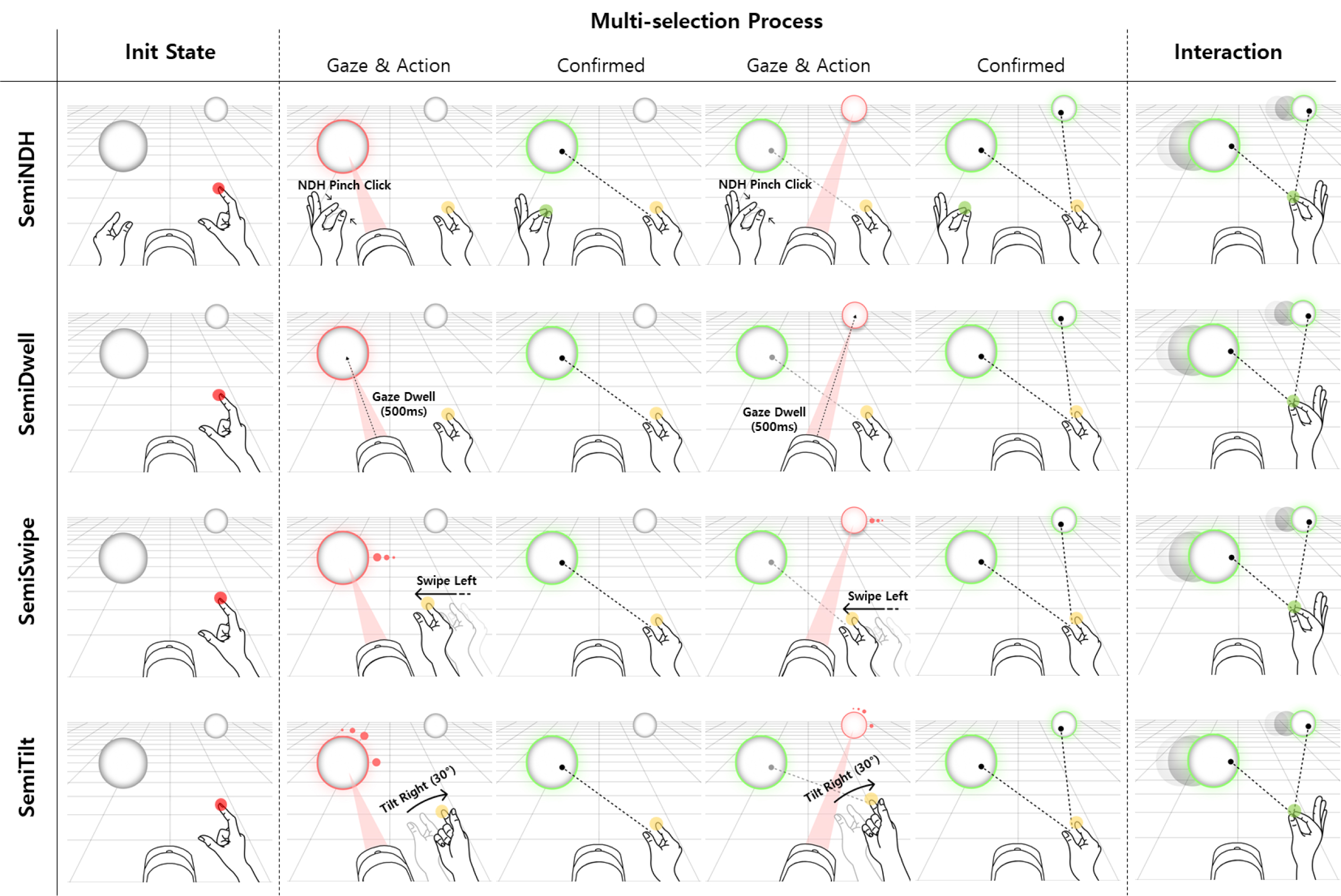}
  \caption{Illustration of \papername, serial multi-selection techniques utilizing the semi-pinch gesture for mode switching. While maintaining the semi-pinch gesture, the user performs a pinch-clicking motion with their non-dominant hand (SemiNDH), maintains gaze for 500 ms (SemiDwell), swipes left (SemiSwipe), or tilts right (SemiTilt) to confirm the grouping of the gazed object. A typical interaction flow involves four steps: (1) maintaining a semi-pinch to retain multi-selection mode, (2) directing gaze at a target and issuing a new command to sub-select it, (3) repeating step 2 until all targets are grouped, and (4) concluding with a full-pinch, which can then be used to drag all targets. After the interaction, users can disable the group by making a full-release pinch.}
  \label{fig:mainInteraction}
  \Description{Illustration of serial multi-selection techniques utilizing the semi-pinch gesture for mode switching. While maintaining the semi-pinch gesture, the user performs a pinch-clicking motion with their non-dominant hand (SemiNDH), maintains gaze for 500 ms (SemiDwell), swipes left (SemiSwipe), or tilts right (SemiTilt) to confirm the grouping of the gazed object. A typical interaction flow involves four steps: (1) maintaining a semi-pinch to retain multi-selection mode, (2) directing gaze at a target and issuing a new command to sub-select it, (3) repeating step 2 until all targets are grouped, and (4) concluding with a full-pinch, which can then be used to drag all targets. After the interaction, users can disable the group by making a full-release pinch.}
\end{figure*}

\subsection{Subselection Triggering Methods}
Next, we propose four subselection triggering methods. These were activated only when the DH was in the semi-pinch state. As in Figure~\ref{fig:mainInteraction}, when an object is gazed at, the outline is highlighted in red (gray in user test) and changed to green (white in user test) if it is grouped. The green outline was disabled when it was ungrouped, and ungrouping individual objects proceeded in the same way.

\subsubsection{Gaze Dwell}
In this method, the gaze cursor had to intersect with the target collider and maintain the activated status (indicated in gray outline as in Figure~\ref{fig:layout} (A)) for 500 ms to group the target~\cite{penkar2012designing, hansen2018fitts, stampe1995selection}. To prevent tracking issues caused by eye-tracking, these invisible colliders were set to have a three times bigger radius than the objects in this study~\cite{lystbaek2024hands}. Also, we tested several variations of gaze duration (250, 500, and 750 ms) and ended up using 500 ms as the best compromise between speed and error. This technique has the advantage of not requiring other hand gestures or movements, which could cause less arm fatigue than other techniques. Although the dwell method could be faster due to raid gaze input, it could potentially be even slower because of the high error rate due to the Midas touch problem reported in the previous studies~\cite{penkar2012designing, isomoto2022interaction}.

\subsubsection{Swipe Left}
Next, we propose a swipe method that uses hand movement to confirm grouping. Inspired by the action in smartphone 
lock screen (slide to unlock)~\cite{arif2014slide}, the user moves their DH toward the left in a semi-pinch state. In this technique, an indicator appears on the right side of the target to show the swipe status, such that users know how much they have to move to confirm grouping. As shown in Figure~\ref{fig:layout} (B), the indicator is designed with three spheres with the bigger ones located closer to the target. The indicator color was displayed white when the target object was in an ungrouped status and gray when it was in a grouped status, indicating the objects' post-activated status. The simple linear movement could enable fast and intuitive subselection. However, given the extensive potential range of hand movements, there is a possibility that they may exceed the intended range of motion.

When using movement-based methods, there was a minor issue of returning to the original position. Since hands can only move a certain distance, a return action is needed to enable a large number of subselections. Thus, we used the left direction of the hand movement and ignored the other side to prevent accidental subselection and make users' return action natural. For implementation, the movement vector of the DH was employed to move the indicator, which exhibited linear movement in a leftward direction from its initial position toward the target. The targets' grouping status was changed when the indicator touched the targets. On the other hand, the utilization of both hand movements on both sides (left and right) may facilitate area selection. This could be achieved by establishing an initial area position for the left swipe action and an end position with the right swipe action. However, this is not addressed in the current paper due to its scope.

\subsubsection{Tilt Right}
Similar to the swipe method, the tilt method also uses a hand movement on the semi-pinch hand. Similarly, as we rotate a knob to adjust the option (i.e., volume, channel) of a radio~\cite{price2021conceptualising}, users can tilt their hand (30\textdegree) to the right to confirm grouping. This technique also provides an indicator for tilt status. The tilt indicator consists of three spheres on the top of the target and has an additional sphere on the right side of the target that indicates the end of the rotation (Figure~\ref{fig:layout} (B)) as we rotate a knob to adjust the volume of radio~\cite{price2021conceptualising}. As the swipe method indicator, the color of the tilt indicator was set to white if the target object was ungrouped and gray when it was in grouped status. In addition, only the right rotation vector was utilized, resulting in a solely rightward indicator movement. To map the hand rotation to the indicator, three times the acceleration was applied to the indicator rotation relative to the actual hand rotation. The target grouping was activated when the top spheres touched the end sphere. Tilt could keep the hand position more static compared to the swipe movement, but there could be potential tracking issues since the gap between fingertips for semi-pinch detection could be occluded during tilting. This false pinch could cause errors when performing a multi-selection (i.e., disable the grouping state by accidental single selection), which could result in increased completion time and fatigue.

\subsubsection{NDH Pinch Clicking}
Pinch clicking uses a full pinch gesture on the NDH to confirm the grouping while maintaining the DH semi-pinch. In this technique, the user maintains the gesture for switching modes and gazes at the target, then performs a full pinch with the other hand to confirm the grouping of objects. The utilization of both hands with a semi-pinch gesture for operation is compatible with the default gesture set, as this particular gesture is not typically employed for other functions. We employ this method for comparisons with the single-handed mode switching techniques and with the movement-based subselection triggering techniques (i.e., swipe and tilt). This could offer advantages in terms of rapid pinch gestures but may require more attention on tracking both hand statuses for mode switching.

\subsection{Interaction after Grouping}
\begin{figure}
  \includegraphics[width=\linewidth]{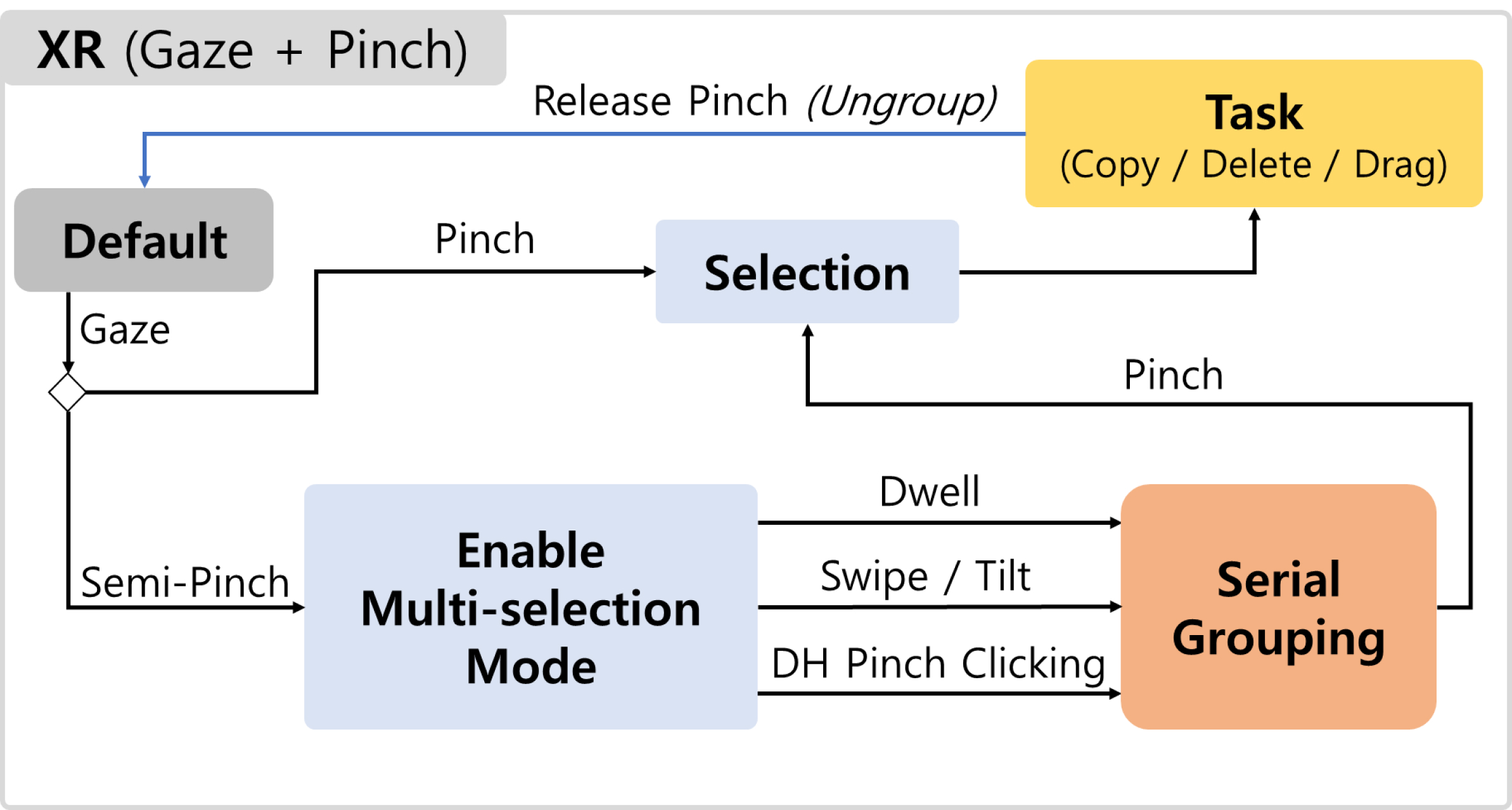}
  \caption{The multi-selection process with \papername.}
  \label{fig:MOSXR}
  \Description{A diagram of a designed multi-selection process in XR environments. Users can utilize the selection function as the default Gaze + Pinch model. In addition, they can perform multi-selection with semi-pinch gestures and trigger subselection with proposed methods.}
\end{figure}

After grouping all targets, users can proceed to the next interaction by making a full pinch as in Figure~\ref{fig:MOSXR}. With the grouped objects, users can perform the same functions on multiple objects simultaneously, such as copy and paste, resize, dragging, etc~\cite{pfeuffer2014gaze, pfeuffer2017gazeP}. Once the interaction has been completed, users could ungroup the targets by making a full-release pinch or add more targets by making a semi-pinch gesture again. 

\section{Evaluation}
We conducted a user study to evaluate the usability and performance of each technique using a multi-selection task of targets on a 2D grid layout in a 3D environment~\cite{bergstrom2021evaluate, wu2023point, lucas2005design}. Since we aimed to design a serial way of multi-selection, we employed the task that resembles selecting a UI component (i.e. photo, file), a common context for grouping under 10 target objects. We measured task completion time, error rate, and amount of hand movement, and collected subjective feedback.

\subsection{Study Design}
\subsubsection{FullDH Technique (Baseline)}
\begin{figure*}
  \includegraphics[width=\textwidth]{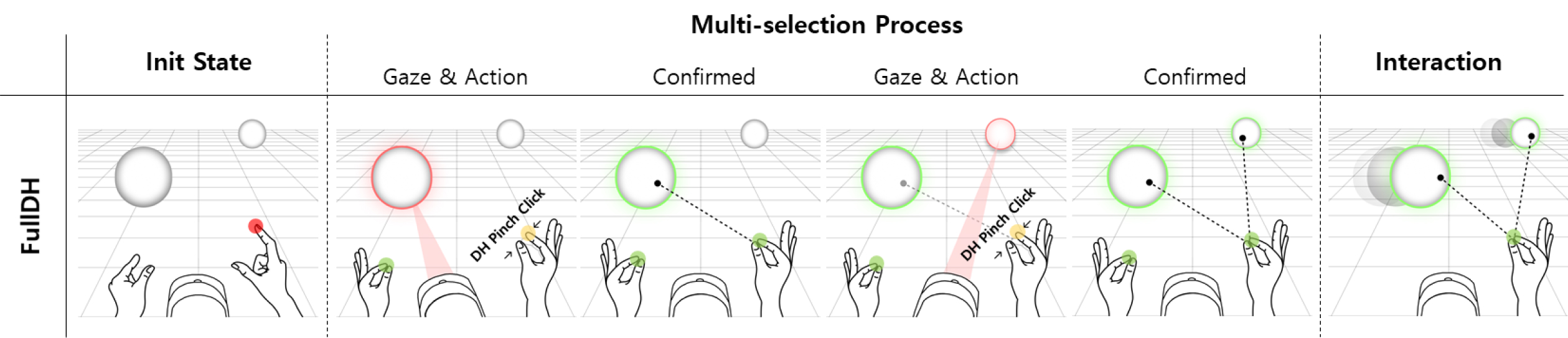}
  \caption{Illustration of FullDH technique process. We included this technique as a baseline, which resembles the mechanism of 2D desktop multi-selection. The FullDH uses a non-dominant hand (NDH) full pinch to activate a multi-selection state instead of a semi-pinch. While maintaining the full pinch gesture, users perform pinch-clicking with their dominant hand (DH) to subselect the gazed object.}
  \label{fig:FullDH}
  \Description{Illustration of FullDH technique process. We included this technique as a baseline, which resembles the mechanism of 2D desktop multi-selection. The FullDH uses a non-dominant hand (NDH) full pinch to activate a multi-selection state instead of a semi-pinch. While maintaining the full pinch gesture, users perform pinch-clicking with their dominant hand (DH) to subselect the gazed object.}
\end{figure*}

We designed a mode-switching method using the NDH full-pinch for comparison with the one-handed semi-pinch method. This technique is based on the metaphor of a 2D desktop environment, pressing the CTRL key and clicking a mouse button for multi-selection~\cite{jarvi2016one, wills1996selection}. As in Figure~\ref{fig:FullDH}, the user first makes a left-hand full pinch to activate the multi-selection mode. While maintaining a left-hand pinch, the user gazes at the target and pinches with their DH for each target serially to make a group. This technique could be intuitive and familiar to users, but it is not compatible with the default gesture interaction model~\cite{pfeuffer2017gazeP}, which uses DH pinch for selection. It requires users to switch the DH and NDH pinch depending on the selection mode.

We have also considered other baselines, such as area selection or UI buttons that enable multi-selection mode. However, these techniques were inappropriate because we were aiming for the serial way of multi-selecting, and buttons are not always accessible in the XR environment. In addition, hypothetically, we could have included a controller baseline, but it would have introduced potential confounds, such as varying tracking quality and haptic feedback, making the study more about a comparison between hand tracking vs. controllers. Thus, considering the scope of this study, we focused on essential factors of gestural design (i.e., uni vs. bimanual, dwell vs. motion vs. pinch) in the eye and hand-tracked XR UIs context. Therefore, we chose the FullDH technique as our baseline as it is similar to holding down the CTRL key in a familiar desktop environment.

\subsubsection{Task Design}
We designed a scattered target grouping task, employed in previous multi-selection studies~\cite{wu2023point, lucas2005design, bergstrom2021evaluate}. The present study did not employ the task based on Fitts' law~\cite{wagner2023fitts, hansen2018fitts}, which is focused on evaluating selection interaction performance. This is because the objective of the present study is to evaluate the proposed techniques in a general context of XR applications and to collect data on overall interaction performance during multi-selection. However, the Fitts' law task focuses on the single selection task and is inappropriate for use in crowded or naturalistic settings.

\begin{figure}
  \includegraphics[width=\columnwidth]{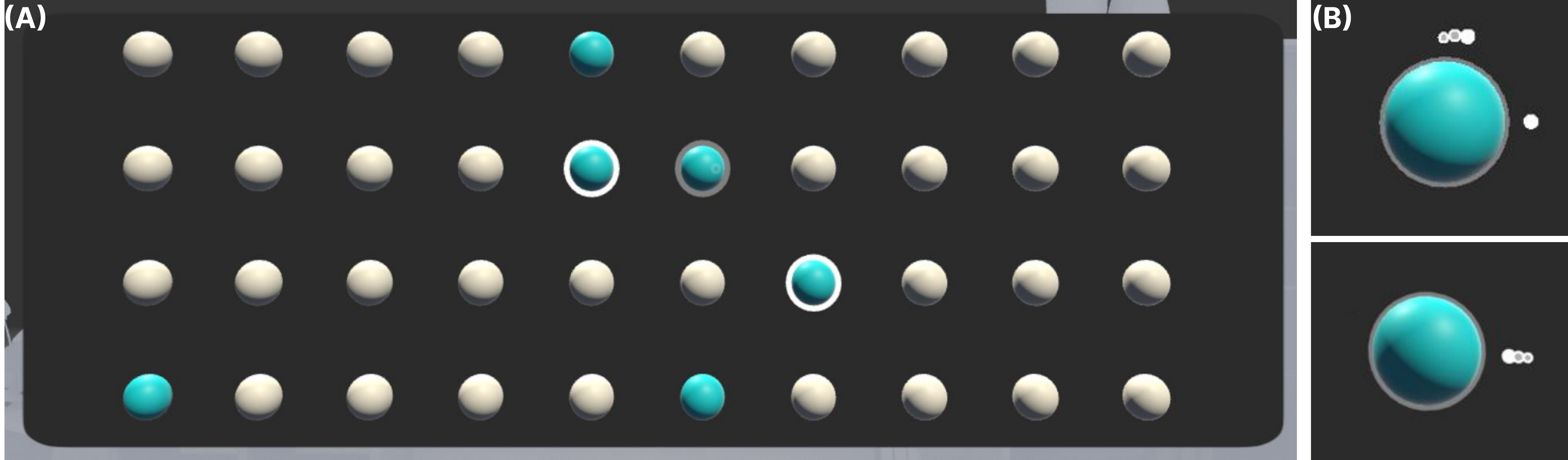}
  \caption{The overall experimental environment and stimulus design. (A) A flat grid layout for object display. A total of 40 objects were placed on the layout, with an equal interval spacing between objects. If the participants gazed at the object, the outline was highlighted in gray. If it was in a grouped state, the object outline was set to white until it was ungrouped. (B) The indicator design for the SemiTilt (top) and SemiSwipe (bottom) techniques, respectively.}
  \label{fig:layout}
  \Description{The overall experimental environment and stimulus design. (A) A flat grid layout for object display. A total of 40 objects were placed on the layout, with an equal interval spacing between objects. If the participants gazed at the object, the outline was highlighted in gray. If it was in a grouped state, the object outline was set to white until it was ungrouped. (B) The indicator design for the SemiTilt (top) and SemiSwipe (bottom) techniques, respectively.}
\end{figure}

A total of 40 sphere-shaped objects were arranged in equal intervals (1m) on a rectangular flat grid layout window (Figure~\ref{fig:layout} (A)) where the objects were randomly placed in each trial. The distance between the participant and the layout window was 13.5 m. The width of the window was set to 58.12\textdegree~, and the height to 25.06\textdegree. The object radius was set to 0.2m each. Participants were asked to group all targets (blue) while avoiding distractors (white).

There were two independent variables in the current within-subject study. First, it consisted of five multi-selection techniques: FullDH, SemiNDH, SemiDwell, SemiSwipe, and SemiTilt. We selected FullDH as a baseline since it is based on the metaphor of a 2D desktop environment using the keyboard and mouse input for multi-selection, which is familiar to most participants. Second, there were three variations in the number of targets: 2, 4, and 6, representing a common range of number of targets for multi-selection~\cite{wu2023point}. We repeated each combination 15 times, resulting in 225 trials. In addition, we included two training trials before each block, consisting of 45 trials in which the grouping task was performed using the specific technique. During the training trial, only 10 spheres were presented in the layout, and participants could practice the technique they would use in the following block. The order of the blocks was counterbalanced using a Balanced Latin square.

\subsubsection{Implementation}
The multi-selection techniques and the study environment were implemented using Unity 3D Engine (2022.3.9f1) and deployed on a Meta Quest Pro (90 Hz, 111.24° FoV) using its embedded eye tracker (30 Hz). We used Meta XR-all-in-One SDK's support for gaze and hand tracking. Due to the minor gaze tracking error of the current hardware system~\cite{hou2024unveiling, sipatchin2020accuracy, stein2021comparison} and layout of 40 small objects, we modified the radius of the invisible object collider, which determines whether the user is gazing at an object, as the previous gaze-based interaction research~\cite{lystbaek2024hands}. We tried to exclude this error to collect data that focused on the multi-selection interaction. Accordingly, the collider was set to three times the radius (60 cm) of the visible object (20 cm). Similar to multi-selection on touchscreen (i.e., touch dwell before dragging grouped files)~\cite{grothaus2011interacting}, we applied pinch-dwell (250 ms) to prevent errors caused by unintended pinch detection due to tracking issues or accidents.

\subsection{Procedure}
\begin{figure}
  \includegraphics[width=\linewidth]{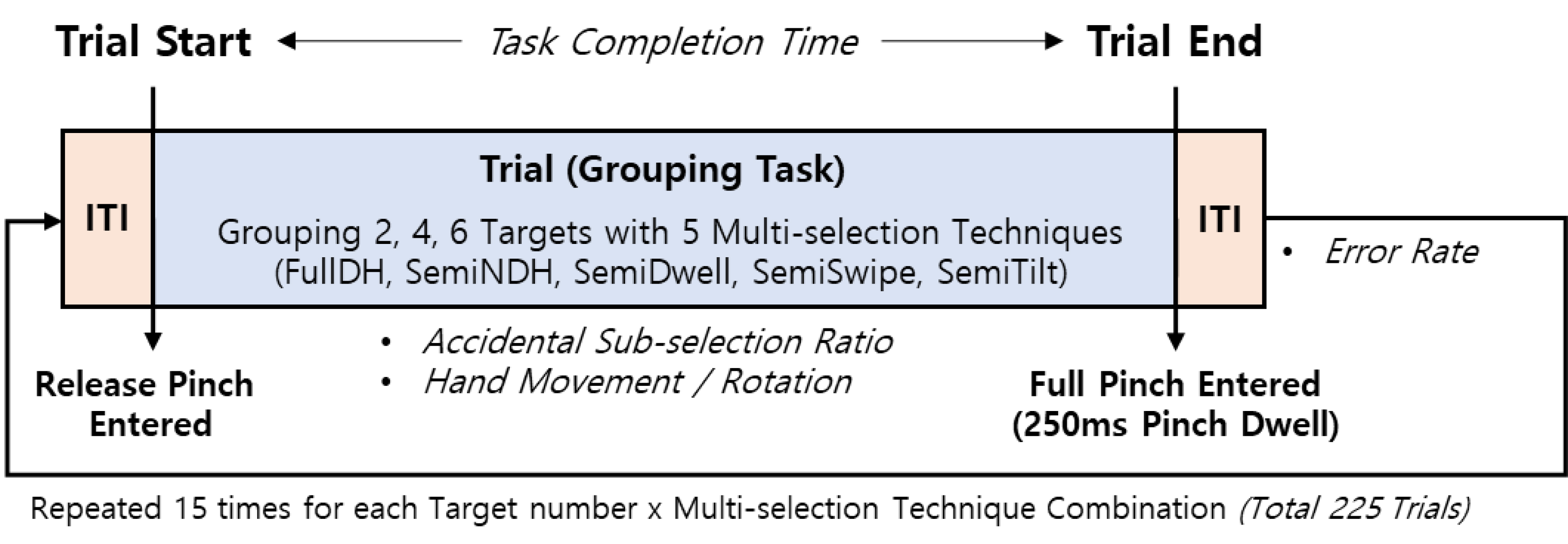}
  \caption{The illustration of the trial procedure. Inter-trial interval (ITI) was provided before and after each trial. The trial started with a release pinch in the ITI state and ended when participants made a full pinch for 250 ms after grouping targets. During the trial, task completion time, accidental subselection case, and hand movement and rotation were collected. After each trial, the error rate was measured by analyzing the grouping result (i.e., the number of targets that failed to be grouped and distractors in the final group).}
  \label{fig:TrialMetric}
  \Description{The illustration of the trial procedure. Inter-trial interval (ITI) was provided before and after each trial. The trial started with a release pinch in the ITI state and ended when participants made a full pinch for 250 ms after grouping targets. During the trial, task completion time, accidental subselection case, and hand movement and rotation were collected. After each trial, the error rate was measured by analyzing the grouping result (i.e., the number of targets that failed to be grouped and distractors in the final group).}
\end{figure}

The experimenter asked participants to complete the consent and demographics forms and briefed them about the task and each multi-selection technique. Then, participants wore the HMD and performed eye-tracking calibration. Before starting each block, two training phases were provided to familiarize the participants with the techniques. Participants spent an average of 24.4 seconds (SD=1.71) and performed 10.77 (SD=3.73) multi-selection in each training phase. During the trial, participants were asked to group all the targets using the assigned technique as fast as possible (Figure~\ref{fig:TrialMetric}). Participants performed and maintained a full pinch for 250 ms after grouping all the targets to complete the trial. After each trial, an inter-trial interval (ITI) scene showing a cross in the center instead of object layouts was presented, and participants were asked to make a full-release pinch gesture (Figure~\ref{fig:mainInteraction}) to proceed to the next trial. After finishing the tasks using each technique, participants were asked to complete a questionnaire about their experience with the multi-selection technique and rest until they were ready to proceed to the next block. After finishing all five blocks, we asked participants to rank the techniques and conducted a brief interview on the ranking. The experiment lasted for 50 minutes on average.

\subsection{Evaluation Metrics}
\begin{itemize}
  \item \textit{Task Completion Time (TCT)}: The interval between the moment when the targets appeared to the moment when a full pinch was performed to end the trial.

  \item \textit{Accidental Subselection Ratio (During Trial)}: The ratio was calculated as the number of distractors (white spheres) grouped accidentally during each trial divided by the total number of subselections performed per trial.

  \item \textit{Error Rate (After Trial)}: The sum of errors after each trial (the number of targets that failed to be grouped by participants, and the number of distractors that were in the final group) divided by the total number of grouped objects per trial.
  
  \item \textit{Inverse Efficiency (After Trial)}: The grouping success rate was calculated as the percentage of trials with no errors (distractor included or target missed to group) in the final group. Inverse efficiency (IE) was calculated by dividing the TCT by the grouping success rate and it indicates the combined effects on grouping efficiency~\cite{statsenko2020applying, townsend1983stochastic}. Therefore, high IE values indicate low grouping efficiency of techniques.

  \item \textit{Hand Movement and Rotation}: The hand movement and rotation while performing multi-selection during each trial were measured. As in previous studies~\cite{bergstrom2021evaluate, wagner2023fitts}, we used the center of the palm translation and rotation value of each frame.

  \item \textit{Questionnaire}: We measured the experience using the System Usability Scale (SUS)~\cite{brooke2013sus}, NASA Task Load Index (NASA-TLX)~\cite{hart2006nasa}, and a single question about satisfaction~\cite{yu2020fully} on a 7-point Likert scale (0 to 6). The SUS is comprised of three positive and negative statements, with the score for the negative statement subtracted from six and subsequently summed with the positive statement scores. The NASA-TLX has six features including mental, physical, and temporal demand, as well as perceived performance, effort, and frustration.

  \item \textit{Ranking \& Brief Interview}: At the end of the study, participants were asked to rank the five techniques in a survey. Open interviews were also conducted regarding the reason for their ranking and their experience with each technique.
\end{itemize}

\subsection{Participants}
We recruited 30 participants (Mean age=25.93, SD=4.63, 15 Male) from the local university. Among the participants, 29 were right-handed, 8 wore glasses, 12 wore contact lenses, and 7 had vision correction surgery. We asked participants about their prior experience of using VR, controller, hand, and eye interaction on a 6-point Likert scale (0-5 points). The participants responded to their VR experience with an average rating of 2.76 (SD=1.41). For interaction modalities, they rated their experience with VR controllers at 2.83 (SD=1.21), with hands at 2.21 (SD=1.61), and with gaze at 2.07 (SD=1.46). All study protocols and methods were approved by the university's Institutional Review Board (IRB), and all participants were rewarded \$10 for participating in the experiment.


\section{Result}
A total of 28385 multi-selections including accidental subselections were collected and it was analyzed using two-way repeated measures ANOVA with five multi-selection techniques (FullDH, SemiNDH, SemiDwell, SemiSwipe, SemiTilt) and three target numbers (2, 4, and 6) as within-subject factors with Bonferroni correction. The Wilcoxon Signed-Rank Tests were used for post-hoc on the distractor-grouped case and error rate metric, due to violations of the normality assumption. For the nonparametric questionnaire data, we used the Friedman test and Wilcoxon signed rank tests for post-hoc analysis. In terms of TCT, hand movement, and rotation analysis, we excluded trials in which participants accidentally skipped or failed to complete the task. Given that the variation in target numbers was even and that the minimum target was two, we filtered out trials where participants failed to group over 50\% of the total targets, resulting in the exclusion of 30 trials (0.44\%).

\begin{figure}
  \includegraphics[width=\linewidth]{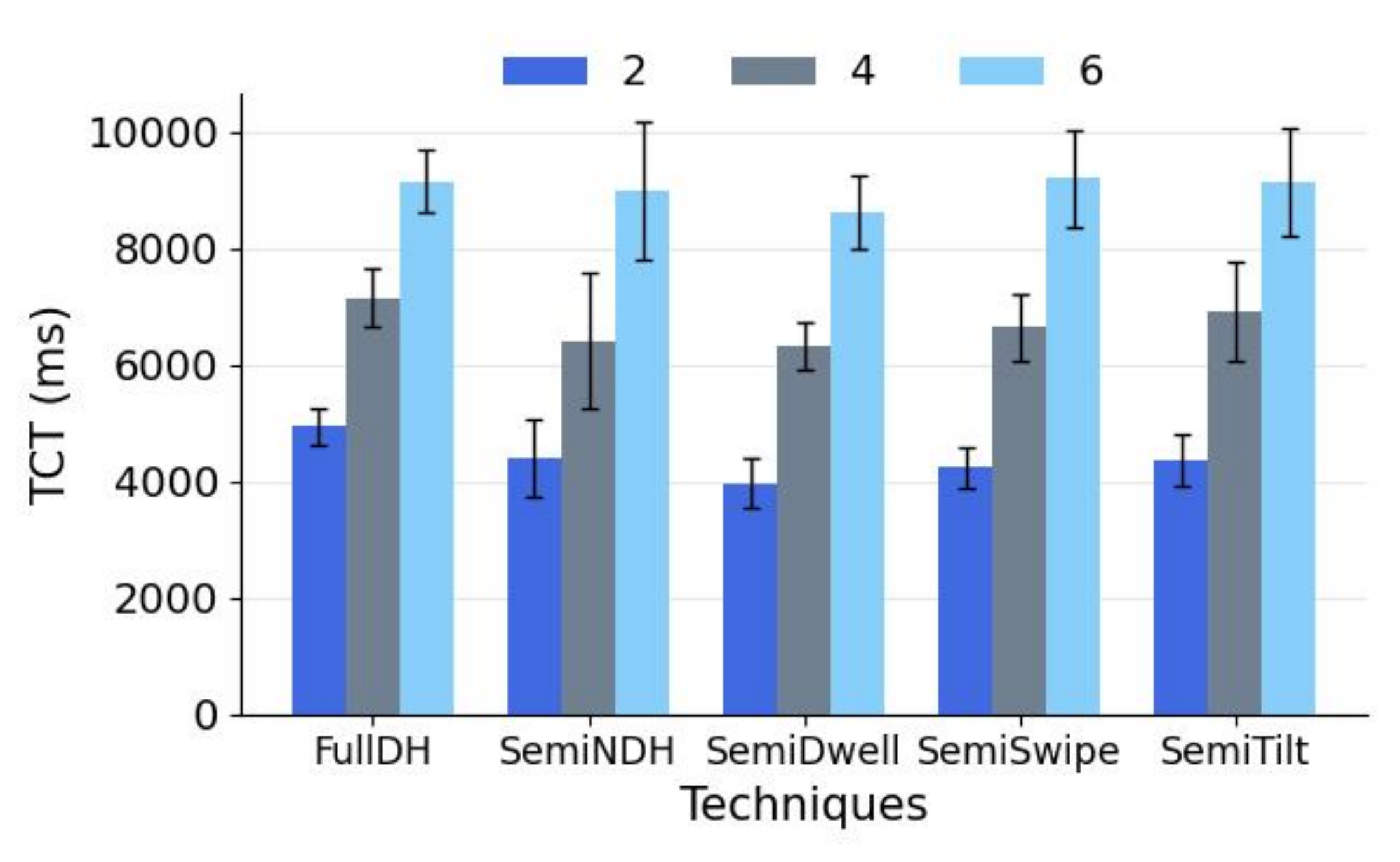}
  \caption{The average task completion time (TCT) in each target number condition and multi-selection technique. Whereas larger target numbers led to longer TCT, no differences were found between the techniques. The error bars represent 95\% confidence intervals.}
  \label{fig:TCTnAcc}
  \Description{The average task completion time (TCT) in each target number condition and multi-selection technique. Whereas larger target numbers led to longer TCT, no differences were found between the techniques. The error bars represent 95\% confidence intervals.}
\end{figure}

\subsection{Task Completion Time (Figure~\ref{fig:TCTnAcc})}
As the number of targets increased ($F_{c}$(1.747, 50.673)=725.511, p<.001, $\eta_{p}^{2}=.962$), participants took more time to complete the task (2 targets (M=4381.942, SD=1293.517) < 4 targets (M=6691.875, SD=2107.008) < 6 targets (M=9029.705, SD=2359.307)) (t(29)s>19.244, ps<.001). However, there was no main effect of multi-selection techniques ($F_{c}$(2.705, 78.457)=1.028, p=.379, $\eta_{p}^{2}=.034$) or the interaction effect ($F_{c}$(5.274, 152.945)=.898, p=.488, $\eta_{p}^{2}=.030$).

We performed the TCT analysis with error-free data by excluding 445 trials (6.5\%). Overall, the test showed similar results, indicating a main effect on target number ($F_{c}$(1.662, 48.192)=749.745, p<.001, $\eta_{p}^{2}=.963$). As the number of targets increased, participants spent more time completing the task (2 targets (M=4382.346, SD=1312.669) < 4 targets (M=6656.641, SD=1989.346) < 6 targets (M=9044.002, SD=2415.898)) (t(29)s>20.398, ps<.001). There was no main effect of techniques ($F_{c}$(2.754, 79.873)=1.276, p=.288, $\eta_{p}^{2}=.042$) or interaction effect ($F_{c}$(4.794, 139.036)=.846, p=.516, $\eta_{p}^{2}=.028$).

\subsection{Accidental Subselection Ratio--During Trial (Figure~\ref{fig:ErrorsnIE} (A))}
The repeated measures ANOVA and post-hoc Wilcoxon signed-rank tests on multi-selection technique ($F_{c}$(3.126, 90.641)=9.550, p<.001, $\eta_{p}^{2}=.248$) indicated that the FullDH (M=8.650, SD=7.633) technique induced a significantly higher accidental subselection ratio than other techniques (Zs>2.746, ps<.006): SemiNDH (M=5.163, SD=4.984), SemiDwell (M=4.206, SD=5.032), SemiTilt (M=3.380, SD=3.177), and SemiSwipe (M=2.436, SD=3.786). In addition, the result showed that the SemiSwipe technique induced significantly less accidental subselection ratio than SemiDwell (Z=2.067, p=.039) and SemiNDH (Z=2.552, p=.011). The main effect of target number ($F_{c}$(1.740, 50.469)=3.150, p=.058, $\eta_{p}^{2}=.098$) and the interaction effect ($F_{c}$(4.980, 144.427)=.760, p=.580, $\eta_{p}^{2}=.026$) were not significant.

\begin{figure*}
  \includegraphics[width=.8\textwidth]{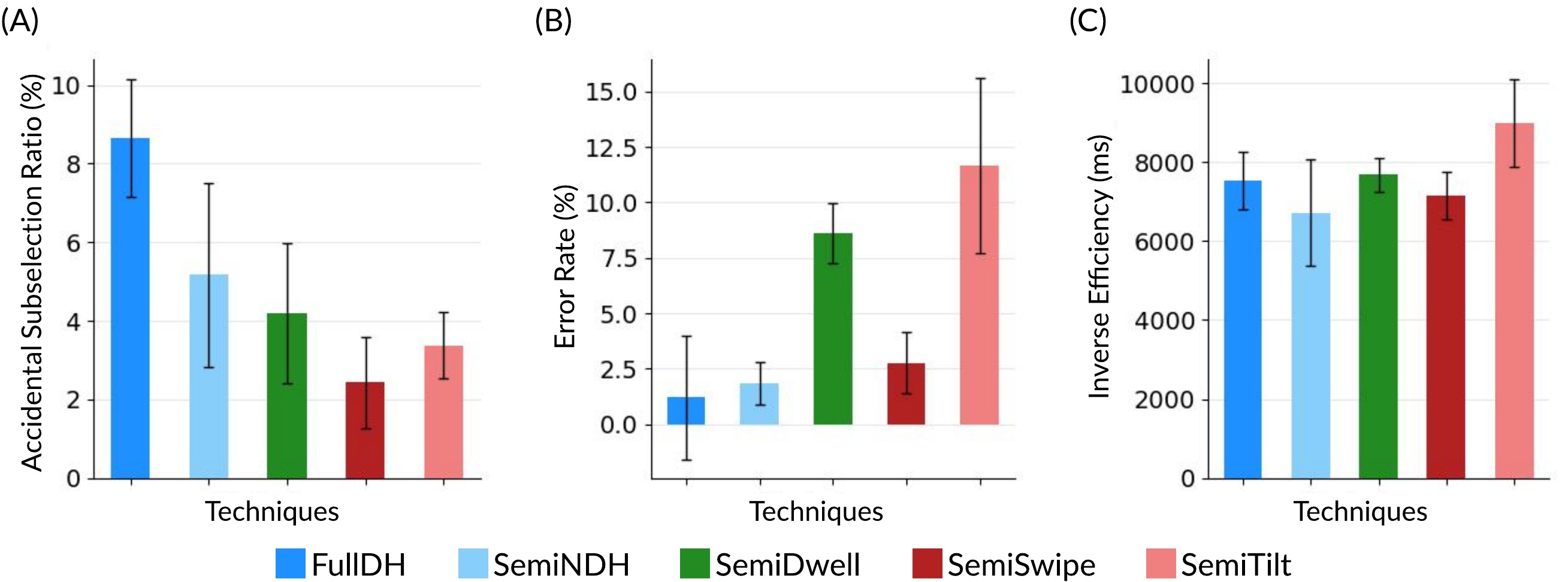}
  \caption{(A) Accidental Subselection Ratio: how often distractors were accidentally subselected out of all subselections during a trial. FullDH showed significantly higher accidental subselection than other techniques. (B) Error Rate: total errors including ungrouped targets and mis-grouped distractors, divided by the total number of grouped objects per trial. SemiDwell and SemiSwipe showed significantly higher error rates than other techniques, but not between them. (C) Inverse Efficiency (IE): task completion time divided by the percentage of error-free trials, with a lower IE indicating better grouping efficiency. SemiNDH was significantly more efficient than SemiTilt. Error bars represent the 95\% confidence intervals.}
  \label{fig:ErrorsnIE}
  \Description{(A) Accidental Subselection Ratio: how often distractors were accidentally subselected out of all subselections during a trial. FullDH showed significantly higher accidental subselection than other techniques. (B) Error Rate: total errors including ungrouped targets and mis-grouped distractors, divided by the total number of grouped objects per trial. SemiDwell and SemiSwipe showed significantly higher error rates than other techniques, but not between them. (C) Inverse Efficiency (IE): task completion time divided by the percentage of error-free trials, with a lower IE indicating better grouping efficiency. SemiNDH was significantly more efficient than SemiTilt. Error bars represent the 95\% confidence intervals.}
\end{figure*}

\subsection{Error Rate--After Trial (Figure~\ref{fig:ErrorsnIE} (B))}
For the main effect of the multi-selection technique on error rate  ($F_{c}$(2.208, 64.023)=19.177, p<.001, $\eta_{p}^{2}=.398$),  SemiSwipe (M=2.767, SD=5.928) showed a significantly lower error rate than SemiDwell (M=8.612, SD=11.483) (Z=4.088, p<.001) and SemiTilt (M=11.648, SD=15.328) (Z=4.165, p<.001), but a higher error rate than FullDH (M=1.207, SD=3.599) (Z=2.974, p=.003). Also, it showed that the SemiDwell and SemiTilt techniques had significantly higher error rates than FullDH (SemiDwell: Z=4.704, p<.001 / SemiTilt: Z=4.535, p<.001) and SemiNDH (M=1.854, SD=6.259) (SemiDwell: Z=4.022, p<.001 / SemiTilt: Z=4.371, p<.001), respectively. There was no main effect of target number ($F_{c}$(1.756, 50.912)=1.710, p=.194, $\eta_{p}^{2}=.056$) or interaction effect ($F_{c}$(3.211, 93.112)=1.996, p=.116, $\eta_{p}^{2}=.064$).

\subsection{Inverse Efficiency--After Trial (Figure~\ref{fig:ErrorsnIE} (C))}
The IE data showed a significant main effect of the multi-selection technique ($F_{c}$(2.606, 75.570)=4.379, p<.001, $\eta_{p}^{2}=.131$). The post-hoc tests showed that the mean IE of the SemiNDH (M=6719.815, SD=1186.851) was significantly lower than the SemiTilt (M=8994.113, SD=3095.398) technique, indicating that the SemiNDH has higher grouping efficiency than SemiTilt. There was no significant difference in efficiency between the other techniques. There was also a significant main effect of target number ($F_{c}$(1.775, 51.469)=274.505, p=.009, $\eta_{p}^{2}=.904$), with the IE increasing as the target number increased (2 targets (M=4795.791, SD=1630.533) < 4 targets (M=7556.147, SD=3107.372) < 6 targets (M=10495.514, SD=3784.793)). Lastly, we found an interaction effect between the technique and target number factor ($F_{c}$(4.916, 142.551)=3.422, p=.006, $\eta_{p}^{2}=.106$). The post-hoc tests showed that the benefit in efficiency for semiNDH over semiTilt was significant starting from grouping 4 target objects ($\Delta$=2243.428, t(29)=3.408, p=.002) and the size of the benefit became even greater when grouping  6 target objects ($\Delta$=3583.793, t(29)=4.686, p<.001).

\subsection{Hand Movement (Figure~\ref{fig:handMove} (A))}
There was a main effect of the multi-selection technique ($F_{c}$(1.339, 38.827)=128.688, p<.001, $\eta_{p}^{2}=.816$) showing that the hand movement of each technique was significantly different from each other (t(29)s>5.542, ps<.001), except between SemiNDH and SemiDwell (t(29)=.677, p=.504). This result indicates that participants had to move their hand most in the SemiSwipe technique (M=1.470, SD=.864), followed by SemiTilt (M=.907, SD=.443) and FullDH (M=.275, SD=.158). Participants moved their hands least when using the SemiNDH (M=.152, SD=.085) and SemiDwell (M=.143, SD=.059) techniques. In addition, there was also a significant main effect of target number ($F_{c}$(1.268, 36.780)=261.889, p<.001, $\eta_{p}^{2}=.900$). Participants moved their hands more as the target number increased (2 targets (M=.354, SD=.333) < 4 targets (M=.593, SD=.628) < 6 targets (M=.822, SD=.892)) (t(29)s>14.034, ps<.001). In addition, there was also an interaction effect between the multi-selection technique and target number ($F_{c}$(2.060, 59.745)=99.846, p<.001, $\eta_{p}^{2}=.775$). The post-hoc results demonstrated that all techniques exhibited significant differences in all target number cases (t(29)s>4.237, ps<.001), except for the comparison between SemiNDH and SemiDwell in all target number cases (t(29)s<.938, ps>.356). These results suggest that all techniques demanded significantly more hand movement as the target number increased. However, the SemiSwipe, SemiTilt, and FullDH techniques exhibited markedly more extreme increases as the target number increased than the other techniques.

\begin{figure}
  \includegraphics[width=\columnwidth]{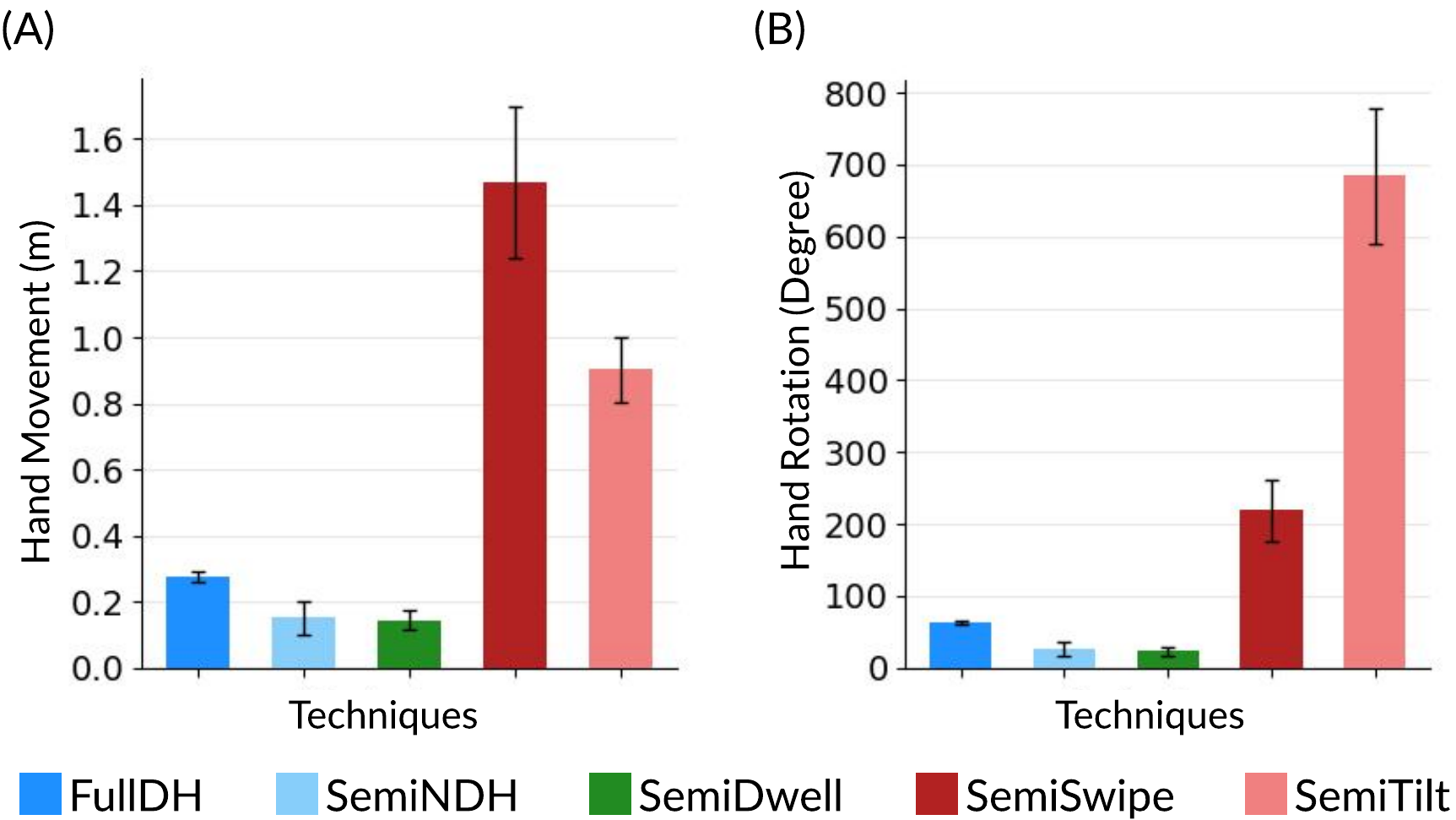}
  \caption{Average (A) hand movement and (B) hand rotation during the trial in each multi-selection technique. All differences between multi-selection techniques were significant except between SemiNDH and SemiDwell in both features, showing they resulted in the least physical hand motion. The error bars represent 95\% confidence intervals.}
  \label{fig:handMove}
  \Description{The average of (A) hand movement and (B) hand rotation during the trial in each multi-selection technique condition. Across (A) and (B), all differences between multi-selection techniques were significant except between SemiNDH and SemiDwell in both features, showing they resulted in the least physical hand motion. The error bars represent 95\% confidence intervals.}
\end{figure}

\subsection{Hand Rotation (Figure~\ref{fig:handMove} (B))}
We found a main effect of the technique ($F_{c}$(1.352, 39.199)=155.639, p<.001, $\eta_{p}^{2}=.843$) indicating that the amount of hand rotation performed during the trial was significantly different from each other (t(29)s>7.833, ps<.001), except between SemiNDH and SemiDwell (t(29)=1.652, p=.109). The SemiTilt technique (M=687.451, SD=389.389) made participants rotate their hands most, and then SemiSwipe (M=219.246, SD=144.122) and FullDH (M=63.040, SD=31.155) followed. In the SemiNDH (M=26.903, SD=16.990) and SemiDwell (M=23.167, SD=10.555) technique conditions, the participants demonstrated the lowest level of hand rotation. There was also a main effect of target number ($F_{c}$(1.404, 40.714)=211.932, p<.001, $\eta_{p}^{2}=.880$) indicating that participants rotated their hands more as the target number increased (2 targets (M=155.715, SD=145.56) < 4 targets (M=208.130, SD=301.49) < 6 targets (M=288.039, SD=410.79)) (t(29)s>11.225, ps<.001). There was also an interaction effect ($F_{c}$(1.870, 54.234)=113.369, p<.001, $\eta_{p}^{2}=.796$) between the multi-selection technique and the number of targets. It indicated that participants rotated their hands more as the target number increased using the SemiTilt, SemiSwipe, and FullDH techniques (t(29)s>6.743, ps<.001). However, there was no significant difference in hand rotation when using the SemiDwell ($\Delta$=2.83, t(29)=1.985, p=.057) and SemiNDH ($\Delta$=.329, t(29)=.312, p=.758) techniques between 4 and 6 targets. SemiDwell only showed a significant difference between 2 and 6 targets ($\Delta$=4.184, t(29)=3.718, p<.001), and SemiNDH showed a significant difference between 2 and 4 targets ($\Delta$=3.721, t(29)=3.553, p<.001) and 2 and 6 targets ($\Delta$=4.050, t(29)=3.689, p<.001).

\begin{figure*}
  \includegraphics[width=\textwidth]{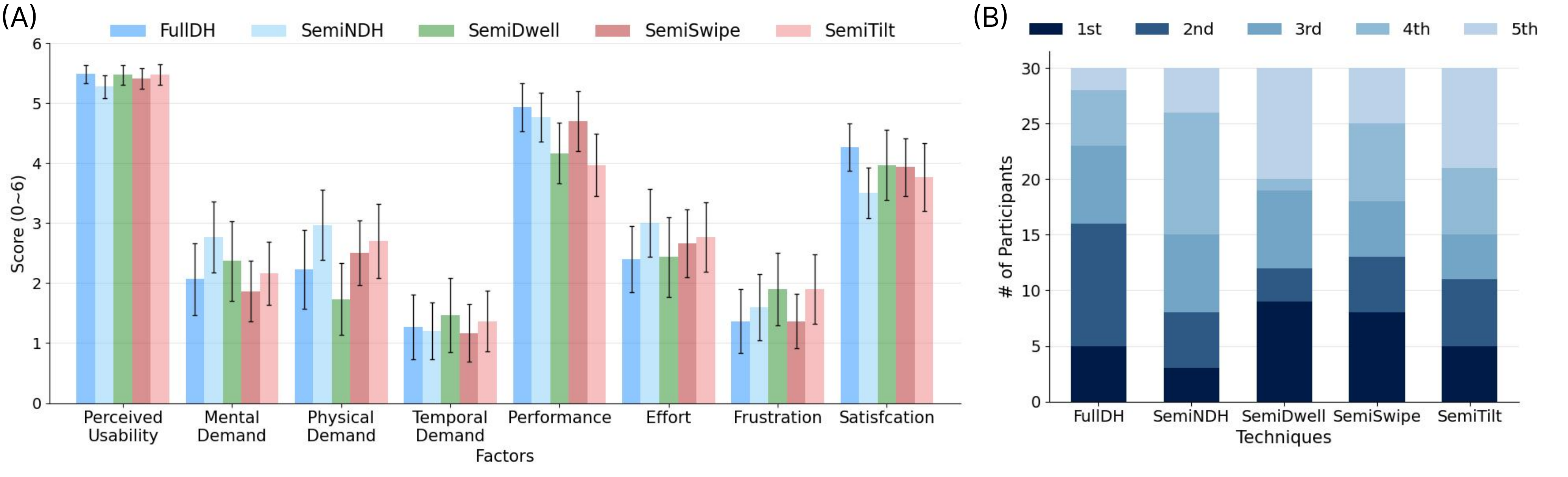}
  \caption{The result of quantitative measures on the experience of using each multi-selection technique. (A) The average score of each factor in the questionnaire (i.e., SUS, NASA-TLX, Satisfaction) on a 7-point Likert scale (0-6). The statistical results indicate that there was a significant main effect of the multi-selection technique on physical demand and performance factors. Also, all techniques have received an average SUS score over 4.05, which is the criterion for minimum usability~\cite{sauro2011practical}. The error bars represent the 95\% confidence intervals. (B) The result of the participants' ranking of each technique.}
  \label{fig:UXQ}
  \Description{The result of quantitative measures on the experience of using each multi-selection technique. (A) The average score of each factor in the questionnaire (i.e., SUS, NASA-TLX, Satisfaction) on a 7-point Likert scale (0-6). The statistical results indicate that there was a significant main effect of the multi-selection technique on physical demand and performance factors. Also, all techniques have received an average SUS score over 4.05, which is the criterion for minimum usability~\cite{sauro2011practical}. The error bars represent the 95\% confidence intervals. (B) The result of the participants' ranking of each technique.}
\end{figure*}

\subsection{Questionnaire (Figure~\ref{fig:UXQ} (A))}
There was a main effect of the multi-selection technique only on physical demand (${\chi}^2(4)$=15.909, p=.003) and performance (${\chi}^2(4)$=20.358, p<.001) for NASA-TLX. The post-hoc result showed that SemiNDH required more physical demand than SemiDwell (Z=3.666, p<.001) and FullDH (Z=2.376, p=.017). Also, the score of SemiTilt was higher than SemiDwell (Z=2.902, p=.004) for physical demand. In terms of perceived performance, both FullDH and SemiNDH demonstrated a higher score than SemiTilt (FullDH: Z=2.757, p=.006 / SemiNDH: Z=2.421, p=.015) and SemiDwell (FullDH: Z=3.111, p=.002 / SemiNDH: Z=1.975, p=.048), respectively. Also, SemiSwipe showed better performance than SemiTilt (Z=2.445, p=.014). Moreover, the perceived usability (SUS) scores of all five techniques averaged over 4.05 points, which is the SUS criterion for minimum usability~\cite{sauro2011practical}.

\subsection{Ranking \& Qualitative Feedback}
Participants responded that they liked the SemiDwell (9) and SemiSwipe (8) techniques the most. However, SemiDwell was also rated as the least preferred technique (10), indicating extreme individual differences in preference. Additionally, 11 participants rated FullDH in second place, while SemiNDH was ranked in fourth place by 11 participants. This shows that most participants also preferred the FullDH but not the SemiNDH technique (Figure~\ref{fig:UXQ} (B)) The detailed rationale for each ranking can be found in the following.

\subsubsection{FullDH}
The majority of participants indicated that the FullDH technique was intuitive and familiar (P7, P19, P20) and fast (P6, P21). P24 mentioned that the sensation of the finger touch when pinching for subselection provided haptic feedback. However, they also indicated that maintaining both hands within the tracking space resulted in greater fatigue than the one-handed multi-selection technique (P3, P13, P22, P29), and it was confusing to keep tracking the states of both hands (P1, P2, P5, P13). Some participants mentioned that since two different functions (group confirm, trial end) were assigned on the right-hand pinch, it occasionally led to errors (P11, P30) due to confusion.

\subsubsection{SemiNDH}
Similar to the FullDH technique, most participants mentioned that the SemiNDH was straightforward (P11, P13, P21) but also caused arm fatigue due to using both hands (P5, P16). Some participants indicated that maintaining the right-hand semi-pinch state required more attention than maintaining the left-hand full pinch in the FullDH technique (P6, P23). Additionally, several participants reported that pinch-clicking with the NDH felt awkward (P15, P24). For instance, P15 mentioned, "I am right-handed, so it felt awkward to pinch with my left hand. I prioritized pinching with my right hand a little higher." In addition, P2 also stated, "It was confusing because I was using my left hand for subselection and then suddenly tried to finish it with my right hand. It felt like I was going back and forth between the two hands."

\subsubsection{SemiDwell}
Participants who preferred the SemiDwell technique mentioned that the absence of hand movement for subselection and one-handed interaction made them physically comfortable (P5, P7, P10, P12, P17). Nevertheless, some have commented that they experienced a lack of stability due to the objects being frequently grouped unintentionally (P5, P8, P9, P11, P27) and experienced significant eye fatigue from repeated usage (P20, P22). Similarly, P4 expressed frustration when attempting to accelerate grouping, given the fixed dwell time of 500 ms.

\subsubsection{SemiSwipe}
Participants indicated that one-handed interaction increased their comfort level (P2, P6, P8, P16). Moreover, participants indicated that SemiSwipe was the most comfortable and stable of the one-handed techniques (P6, P23, P27, P28, P29). They mentioned that SemiSwipe was more stable and faster for multi-selection due to its gaze independence (P2, P10, P16, P28). However, participants indicated they needed time to familiarize themselves with the technique (P24, P25). In addition, they mentioned the need to consider the range of action and how the hand returns (P5, P11). They also reported experiencing physical fatigue due to the additional action required compared to other techniques.

\subsubsection{SemiTilt}
Regarding the SemiTilt technique, participants mentioned they could group faster using a simple one-handed movement (P2, P22, P27). However, similar to the SemiSwipe technique, participants also commented on physical fatigue, with more tension required on the wrist and shoulder in the SemiTilt technique (P5, P6, P14, P17, P19, P23). They also considered the action of returning to the initial hand position ambiguous (P11, P13), and SemiTilt more sensitive to hand movement (P9, P25, P28, P30).


\section{Discussion}
\subsection{Semi-pinch based One-handed Multi-selection Interaction}
Regarding the FullDH technique, the keyboard-mouse metaphor would have been beneficial because of the familiarity with the habit and motor concept of using gaze along with DH. We found this aspect from the interview and questionnaire results. In contrast, the pinch-clicking action with the NDH in the SemiNDH technique made participants feel more confused and awkward. This finding aligns with the bimanual interaction principles proposed by Guiard~\cite{guiard1987asymmetric} that guide the utilization of NDH for coarse, irregular action and DH for detailed and frequent action to activate the designated function. Thus, participants found the FullDH technique, which assigned mode switching to NDH pinch and rapid pinch-clicking for subselection to DH, more familiar and straightforward than the SemiNDH technique.

However, there were limitations regarding the coupling between gaze and pinch gestures. For instance, the familiar and fast mechanism tended to induce more accidental subselection during grouping, and participants had to keep track of both hand states while performing multi-selection. These limitations caused too many errors during the trial, which had to be corrected afterward. Eventually, the bimanual techniques could perform fast subselection but resulted in TCT similar to that of other one-handed techniques. Still, as the previous work of bimanual interaction indicates that the performance could differ depending on the tasks~\cite{lystbaek2024hands, hinckley1997cooperative}, the FullDH technique could be affordable outside of contexts that require quick and accurate subselection.

Overall, the semi-pinch-based one-handed techniques showed better performance in error during subselection and user evaluation, especially when a swipe was used for subselection. Participants preferred the one-handed technique since it required less attention on the hand gesture status and enabled them to focus solely on the grouping task. Also, the semi-pinch provided an intuitive and straightforward interaction process for participants, such that they could adapt quickly. However, there are also limitations in semi-pinch-based techniques, such as the varying performance depending on the subselection triggering method.

With the one-handed technique, users can perform other tasks (e.g., moving objects before sub-selecting for alignment, resolving occlusion, scrolling down lists) that could support multi-selection, whereas the bimanual technique requires all resources for one task. For instance, users can remove an occluding object with their NDH and then perform multi-selection with their DH. Therefore, the one-handed multi-selection technique has the potential to enhance compatibility with other pinch-based interactions, not only interaction after grouping but also throughout the multi-selection process.

\subsection{Hand Movement for Triggering Subselection}
Compared to the pinch clicking and dwell method for subselection, we found that the method with hand swiping outperformed other methods in terms of error and usability. Participants also mentioned that gaze dwell was comfortable since it required no extra hand movement. However, the error metric results indicated that the Midas touch problem limited the gaze-dependent technique (SemiDwell). Thus, to adopt the dwell-based technique in the application, appropriate dwell time lengths need to be adjusted for each user with the value that they can tolerate and handle the errors.

Pinch-clicking-based methods (i.e., FullDH and SemiNDH) caused high errors due to gaze-independent aspects (i.e., no visual indicators and sudden/rapid subselection); in contrast to the hand movement-based techniques, the visual features, such as an indicator of gaze and hand status, are less involved in the subselection process. The pinch action could be sudden and fast, making it difficult for users to track their selection status while sub-selecting. As a result, users could occasionally make the pinch gesture before their gaze touches the object they want to sub-select and make an early trigger error~\cite{kumar2008improving}. Users could redo the selection if they selected the wrong object in a single selection task, but it could be problematic in a multiple selection task that requires users to correct their mistakes additionally.

The current study utilized micro-gestures such as hand swipe and tilt for the subselection method. However, the compatibility of the micro-gesture with gaze must be considered when designing a subselection triggering method~\cite{pfeuffer2024design}. We found that the visual indicator plays an important role in enhancing this compatibility. For instance, the swipe method showed an overall low error rate and comparable performance to the pinch clicking-based techniques in the NASA-TLX outcome. The linear swipe and 1:1 indicator to hand movements were intuitive and worked positively for the participants. Thus, the results of this study indicated that SemiSwipe is the most appropriate technique for multi-selection in an eye-hand interaction context among the designed techniques.

On the contrary, participants did not prefer the SemiTilt technique in this study, not only because it caused more arm fatigue but also because of the high sensitivity of indicator movement to
hand movement. We tried to prevent potential tracking issues, such as false pinch detection when rotating DH, by employing the threshold of 30\textdegree~ instead of 90\textdegree. Still, several participants experienced tracking errors during the trial and felt it was too sensitive. This could depend on the participants' pinch style (i.e., preferred pinch gesture direction) and hand size. The incorporation of sophisticated hand-tracking capabilities for micro-gestures~\cite{pei2024ui, kin2024stmg} and individual hand calibration for personalized semi-pinch detection threshold~\cite{zhu2023pinchlens}, have the potential to enhance the precision and usability of SemiTilt.


\section{Applications}

\begin{figure*}
  \includegraphics[width=\textwidth]{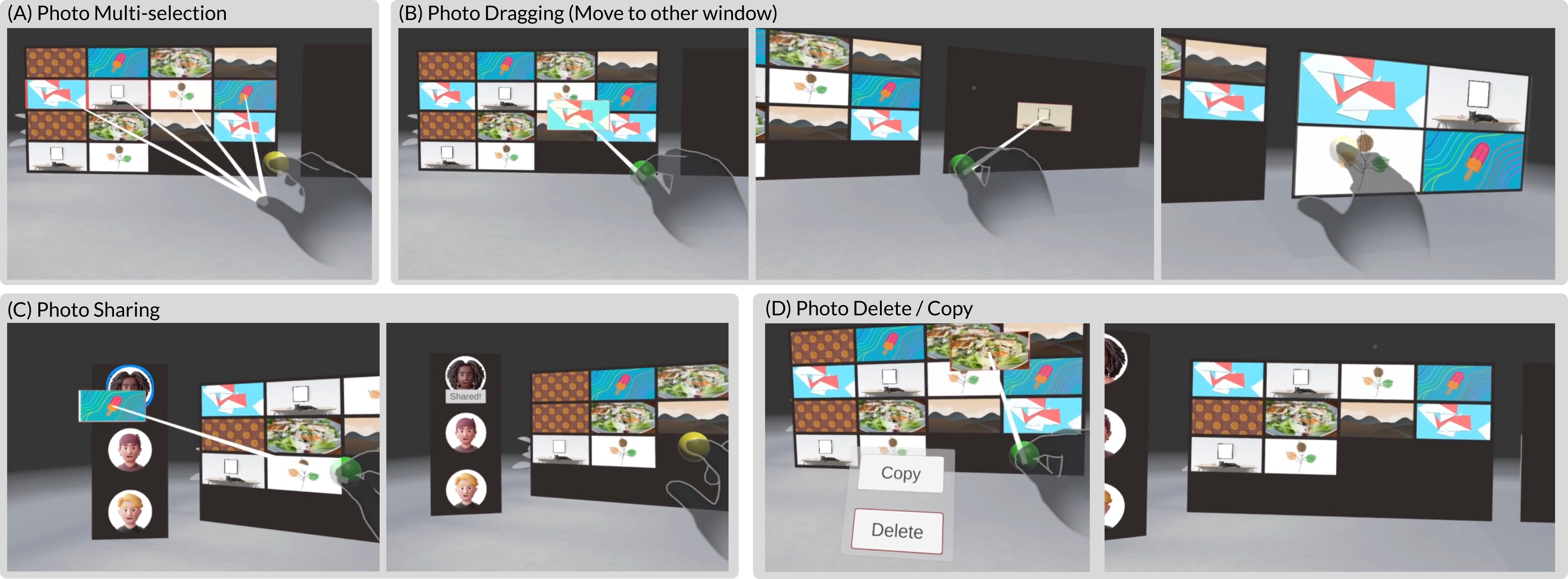}
  \caption{The file management application: (A) Users can group multiple photos with the \papername techniques. (B) Drag and relocate photos to other windows/folders. (C) Share photos with other users. (D) Selecting an option for copying or deleting the grouped photos with gaze and full-release pinch. After the function was activated, the grouped photos were deleted.}
  \label{fig:FileMng}
  \Description{The file management application: (A) Users can group multiple photos with the \papername techniques. (B) Drag and relocate photos to other windows/folders. (C) Share photos with other users. (D) Selecting an option for copying or deleting the grouped photos with gaze and full-release pinch. After the function was activated, the grouped photos were deleted.}
\end{figure*}

\begin{figure*}
  \includegraphics[width=\textwidth]{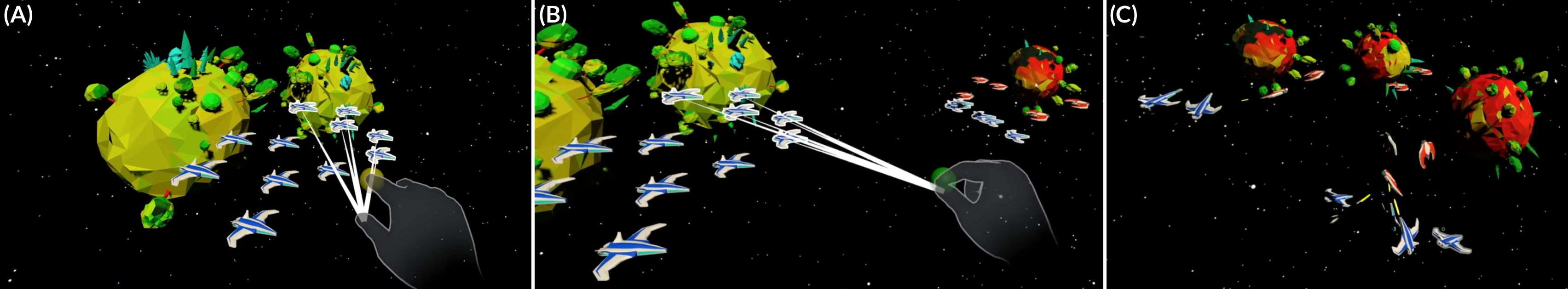}
  \caption{The RTS game application: (A) Grouping fighters by using the SemiSwipe technique, (B) Dragging multiple fighters to appropriate locations based on their strategy, (C) Start combat with the enemy fighters.}
  \label{fig:Game}
  \Description{The RTS game application: (A) Grouping fighters by using the SemiSwipe technique, (B) Dragging multiple fighters to appropriate locations based on their strategy, (C) Start combat with the enemy fighters.}
\end{figure*}
Based on the study results, we illustrate two scenarios in which multi-selection could be generally used in XR applications: File management and RTS Game. By presenting the application scenarios, we aim to provide direct insight and design guidelines for adopting multi-selection functions. All five techniques are presented in each application scenario with the same interaction process as the user test. However, we applied additional visual representation for grouping status by adding an individual line connecting grouped objects to the right-hand fingertip for clearer indication. 

\subsection{2D Environment Context: File Management}
First, we present the file management application, which represents a fundamental task in the 2D environment for various domains (i.e., programming, design, office work)~\cite{jarvi2016one, lucas2005design, dehmeshki2010design}. Since file management mostly consists of sub-tasks such as dragging files to other folders or applying specific operations (i.e., deleting, duplicating, copying, compressing) on multiple files, a multi-selection function would be more useful. Furthermore, providing an efficient method for managing multiple objects in XR becomes more crucial, as recent XR is capable of presenting multiple windows at once~\cite{li2024predicting, cheng2023interactionadapt}.

As in Figure~\ref{fig:FileMng}, we implemented a photo management application with sharing, moving, deleting, and copying functions. Users can select a photo that they want to manage by gazing at it and making a semi-pinch to select multiple photos using the \papername techniques. After selecting the photos, users can share them with other people by dragging them to the target avatar. They can also move the photos to other folder windows in the same way. Next, when users hold a full pinch with the photos selected for 2s, the button interface for the delete and copy option is displayed. Then they can select the deleting and copying function by gazing at the button and releasing a pinch to activate the function.

\subsection{3D Environment Context: RTS Game}
The XR environment consists of more complex features and contexts than the 2D environment, which makes it difficult to respond to the previous multi-selection techniques that require multiple steps for mode switching. However, we have found that our multi-selection design allows for rapid grouping and intuitive mode switching based on user test results. As a result, our techniques are adaptable and responsive to more complex and dynamically changing situations. Additionally, the multi-selection function is capable of maintaining alignment between objects during dragging~\cite{lucas2005design, shi2022group}. This is an advantage in contexts where the 3D models need to be moved in formation, such as furniture placement or real-time strategy (RTS) gaming. Therefore, we present RTS gaming as an example of a 3D environment context that could represent both complex context response and alignment features of multi-selection.

As in Figure~\ref{fig:Game}, we implemented an RTS game in which players have to protect their planets from the invasion of enemy fighters. As in the file management application, the players can select the fighters by gaze and pinch interaction, and perform multi-select with the designed techniques. When the fighter is placed close enough to the enemy, it begins to engage. With the proposed multi-selection technique, players could quickly and appropriately distribute their fighters based on their strategy and also maintain the alignment to attack the enemy in a more efficient formation.


\section{Limitations \& Future work}
The techniques featured in the current study were all designed to perform serial multi-selection tasks, without covering parallel techniques for scenarios such as selecting a large number of targets that are densely arranged~\cite{wu2023point, shi2024experimental}. While we believe that semi-pinch-based gaze selection techniques are most suitable for serial multi-selection tasks that cover the most common scenarios in XR, we concur that parallel multi-selection techniques also need to be investigated in the future. One potential parallel technique based on our proposed methods is using a swipe movement while maintaining a semi-pinch. This technique could allow the user to define two points by swiping left and right, respectively, in the 3D space for area selection.

Since we focused on the Gaze + Pinch interaction context, we compared hand-related features (i.e., unimanual vs. bimanual and dwell vs. motion vs. pinch) within the interaction. However, comparison with other input modalities, such as ray-casting with a controller, could provide additional insights into usability. In addition, the efficiency of each confirmation method could be explored more in the Fitts' law-based task. These were not explored in this study, but these features could strengthen the insight for multi-selection interaction in XR and should be explored in the future.

There were also some restrictions in the implementation variables, such as the collider size for target selection and the threshold distance for pinch state. For the proposed techniques to be universally applicable, it would be essential for these values to be adjustable depending on the context and individual characteristics. Personalization approaches based on the hand size or exploring the usage of left-handed users could help reduce the cognitive load of using semi-pinch gestures.

The current study presents the objects on a 2D flat layout with different colors that differentiate between targets and distractors. However, in most actual applications, targets are not highlighted, necessitating users to search and classify the objects before grouping them. For example, when users want to remove files from the list, they must consider the file features (e.g., name, size, date, etc.) to select the files that need to be removed. Also, there are diverse features in the XR environment, such as depth and dynamic targets, that could affect the performance. In these contexts, gaze-dependent techniques such as SemiDwell may result in more errors due to the Midas Touch problem~\cite{penkar2012designing, isomoto2022interaction}. Consequently, further study is required to explore multi-selection techniques in broader contexts that consider 3D features to suggest appropriate techniques for future XR applications.

Although the SemiSwipe technique was preferred and showed better performance than other techniques, participants tended to swipe more than the activation threshold. The SemiTilt also had a similar but more critical issue, resulting in poor performance and evaluation. These limitations could be improved with the advanced algorithm or additional tracking sensors. As our study demonstrated that visual feedback is a key factor in the compatibility of hand movement with gaze, it may also have the potential to improve these limitations. Therefore, future research could explore various forms of indicators and feedback that could indicate the subselection status and prevent excessive movement. For instance, auditory or tactile feedback could be used with visual feedback to enhance the awareness of users in tracking the subselection status~\cite{jang2024effects}.

\section{Conclusion} 
This paper proposes the \papername, techniques for multi-selection function for eye-hand interaction in the XR environment. We designed one-handed multi-selection techniques using a semi-pinch hand gesture for mode switching while using microgestures for subselection (i.e., SemiSwipe, SemiTilt). In the user study, we compared these techniques with the bimanual dominant-hand full pinch method (i.e., FullDH) and non-movement-based subselection triggering methods (i.e., SemiNDH, SemiDwell). As a result, one-handed interaction with semi-pinch mode switching and swipe hand movement for subselection (SemiSwipe) showed the least errors, received higher scores on the evaluation, and was most preferred by the participants. Based on these findings, we presented application scenarios where our techniques could improve the user experience of multi-selection tasks in XR environments.

\begin{acks}
This research was supported by the National Research Council of Science \& Technology (NST) grant by the Korea government (MSIT) (No. CRC21014) and the STEAM R\&D Project, National Research Foundation (NRF), Republic of Korea (RS-2024-00454458).
\end{acks}
\balance
\interlinepenalty=10000 
\bibliographystyle{ACM-Reference-Format}
\bibliography{CHI25_MOS}
\end{document}